\documentclass[%
 reprint,
superscriptaddress,
 amsmath,amssymb,
 aps,prb,
8pt]{revtex4-1}

\usepackage{tikz}
\usepackage[export]{adjustbox}
\usepackage{graphicx}
\usepackage{dcolumn}
\usepackage{bm}
\usepackage{physics}
\usepackage{float}
\usepackage{subfig}
\usepackage{color}
\usepackage{array}
\usepackage{makecell}
\usepackage{ragged2e}
\usepackage[font=small, justification=centerlast, singlelinecheck=off, textformat = simple, format = plain]{caption}

\usepackage{bbold}
\usepackage{fancyhdr}
\usepackage{CJK}

\begin{document}

\preprint{APS/123-QED}

\title{Possibility to detect the bound state of the Heisenberg ferromagnetic chain at intermediate temperature}

\author{Mithilesh Nayak}
\email{mithilesh.nayak@epfl.ch}
\affiliation{Institute of Physics, Ecole Polytechnique F\'ed\'erale de Lausanne (EPFL), CH-1015 Lausanne, Switzerland}

\author{Fr\'ed\'eric Mila}
\email{frederic.mila@epfl.ch}
\affiliation{Institute of Physics, Ecole Polytechnique F\'ed\'erale de Lausanne (EPFL), CH-1015 Lausanne, Switzerland}

\date{\today}

\begin{abstract}
Motivated by the lack of direct evidence with inelastic neutron scattering of the well documented bound state of Heisenberg ferromagnets,
we use the time-dependent Thermal Density Matrix Renormalization Group algorithm to study the temperature dependence of the dynamical spin structure factor of Heisenberg ferromagnetic spin chains. For spin-1/2, we show that the bound state appears as a well defined excitation with significant spectral weight in the temperature range $J/12 \lesssim T \lesssim J/3$, pointing to the possibility of detecting it with inelastic neutron scattering near $k=\pi$ provided the temperature is neither too low nor too high - at low temperature, the spectral weight only grows as $T^{3/2}$, and at high temperature the bound state peak merges with the two-magnon continuum. For spin-1, the situation is more subtle because the bound state with two neighboring spin flips competes with an anti-bound state with two spin-flips on the same site. As a consequence, the relative spectral weight of the bound state is smaller than for spin-1/2, and a weak resonance due to the anti-bound state appears in the continuum. 
A clearer signature of the bound state (resp. anti-bound state) can be obtained if a negative (resp. positive) biquadratic interaction is present.

\end{abstract}
\maketitle
\section{Introduction}
The first evidence of bound states of spin waves in ferromagnets goes back to 1931 and Bethe's solution of the spin-1/2 Heisenberg chain\cite{Bethe_1931}, only one year after Bloch's theory of spin waves\cite{Bloch}.
Since the Heisenberg ferromagnet conserves the number of spin-deviations and the states with different spin deviations are orthogonal to each other, Bloch realised that the low lying excitations are better understood in terms of spin-deviations, and that the Heisenberg model can be simply diagonalised in the subspace of single spin-deviation states in momentum basis, leading to the concept of spin-waves\cite{Bloch}. Similarly, the model can be studied in the subspace of two spin-deviation states, and the problem can be reduced to a one particle problem in the centre-of-mass momentum basis, with quite generally a continuum of two-magnon excitations and a well separated bound state emerging from two neighboring deviations\cite{Bethe_1931,Dyson_1956, Wortis_1963, DCMattis, Haldane_1982a, Haldane_1982b}. In 1D, the bound state exists as a separate excitation for all values of the wave-vector $k$.

The Heisenberg ferromagnet is realised in nature for example in the spin-1/2 compound $(\mathrm{C}_6\mathrm{D}_{11}\mathrm{N}\mathrm{D}_3)\mathrm{Cu}\mathrm{Br}_3$ or in the spin-1 compounds $\mathrm{CsNi}\mathrm{F}_3$ and $\mathrm{Ni}\mathrm{Nb}_2\mathrm{O}_6$, to cite a few \cite{Mikeska_1977, Bohn_1980, Kopinga_1986, PChauhan_2020, Tapan_Neutron_Scattering}. Inelastic neutron scattering (INS)  experiments have probed spin-wave excitations by measuring the differential scattering cross-section in these ferromagnetic compounds\cite{Kopinga_1986,DeVries_1989}. However, detecting the bound state of spin-waves in a ferromagnet has remained a challenge because INS measures single spin-flip excitations. At low temperatures the thermal ensemble is predominantly populated by fully aligned states, and accordingly one observes spin-wave excitations only. However, upon increasing the temperature, the thermal ensemble is populated with single-spin-deviation states and a spin-flip in these states results in states in the two-spin deviation subspace. Thus, INS experiments are in principle able to capture the bound-state of ferromagnets if they are performed at a non-zero temperature. 

There have been previous attempts to observe the bound states of spin-waves. Silberglitt and Harris have demonstrated that the bound states, in a 3D ferromagnet, have observable signatures in the thermal dynamical structure factor (DSF) in the large wavelength limit ($k=0$)\cite{Silberglit_Harris_1967,Silberglitt_Harris_1968}. The resonance of the bound state with the two-spin-wave continuum results in a broadening of the line width and an increase of the intensity of the single spin-wave peak which, a priori, can be detected in INS experiments. However, although the bound states of 3D ferromagnets exist as separate modes above a certain threshold of $k$, they are expected to gather a very small spectral weight compared to the main spin-wave excitations and to lie too close in energy to be properly resolved after thermal broadening \cite{Silberglitt_Harris_1968}. Thus, the direct detection of bound states in INS experiments is difficult. Instead far-infrared transmission techniques have been used to find indirect signatures, and ferromagnetic resonances have been observed for $\mathrm{Co}\mathrm{Cl}_{2}.2\mathrm{H}_2 \mathrm{O}$(CC2)\cite{Date_Motokawa_1966, Torrance_Tinkham_1969}. However, one cannot completely distinguish the bound state  as the far infrared measurements are close to $k = 0$ where the bound state is not well-resolved from the single spin-wave excitations.

In the case of 1D ferromagnets the difference in excitation energy between the continuum and the bound state is largest at $k=\pi$, and this difference is larger than in its 3D counterpart. Therefore, a finite temperature INS experiment has a better chance to resolve and detect the bound state of spin-waves in 1D. To check this expectation, we directly simulate the DSF using finite temperature time dependent Density Matrix Renormalization Group algorithm (henceforth referred to as thermal t-DMRG algorithm). We find evidence of the bound state and compare its spectral intensity with that of the single spin-wave peak. Although bound states exist for ferromagnets with arbitrary spin, we focus our investigation to the finite temperature dynamics of spin-1/2 and spin-1 ferromagnetic chains.
  
  The paper is organised as follows: in section \ref{method}, we briefly describe the numerical method we used to obtain the results. In section \ref{spin-1/2 FM chain}, we discuss the thermodynamics of the spin-1/2 FM chain obtained from thermal DMRG simulations and benchmark it with Wang-Landau Stochastic Series Expansion Quantum Monte Carlo (QMC) algorithm from ALPS package. We then report on the finite temperature dynamics of spin-waves for the spin-1/2 FM chain. In order to characterise the bound state, we measure the spectral weights associated with it in the isotropic case and give simple arguments to motivate the shape of the spectral peak and the nature of its temperature dependence which we support with spin-wave calculations done in the limit of vanishing magnetic field in section \ref{Thermal_DSF_mag_field}. In section \ref{spin-1 FM chain}, we extend the discussion to spin-1 chains.  We discuss the two-spin deviation spectrum of the spin-1 FM chain using the dynamical quadrupolar structure factor at zero temperature, and we explain the origin of resonances in the two-magnon continuum by including biquadratic interactions. At finite temperature, we find clear signatures of the bound state in the thermal DSF of the Heisenberg model and of anti-bound states for large biquadratic interactions, and we extend the discussion to the case with easy axis anisotropy.

\section{The method}  
\label{method}
INS experiments measure a differential scattering cross-section which is directly proportional to the dynamical structure factor (DSF) denoted as ${S^{\alpha\tilde{\alpha}}(k,\omega)}_{\beta}$. DSF is the Fourier transform of a time-dependent correlation function denoted $C^{\alpha\tilde{\alpha}}(l,t;\beta)$ defined by:
\begin{eqnarray}
&&C^{\alpha\tilde{\alpha}}(l,t;\beta) = \mathrm{Tr}\lbrack \hat{\rho}_{\beta}S^{\alpha}(\Delta r_l;t)S^{\tilde{\alpha}}(0;0)\rbrack\nonumber\\
&&\\
\label{Spin_DSF_def}
&&{S^{\alpha\tilde{\alpha}}(k,\omega)}_{\beta} = \frac{1}{L^2}\int_{-\infty}^{\infty}dt\sum_{l}e^{-i(k\Delta r_l+\omega t)}C^{\alpha\tilde{\alpha}}(l,t;\beta)\nonumber
\end{eqnarray} 
where, $L$ is the number of sites and $\Delta r_l$ is the relative position with respect to the centre of the chain at $l=0$ and $\beta$ is the inverse temperature. The thermal DSF ${S^{\alpha\tilde{\alpha}}(l,t)}_{\beta}$ can represent different components depending on $\alpha,\tilde\alpha\in\lbrace +, - , z\rbrace$.
The computation of the DSF at finite temperature can be performed by using a time-dependent DMRG algorithm on a thermal ensemble (denoted as $\hat{\rho}_\beta$) pioneered by Barthel et al \cite{Barthel_2009, Barthel_2013} and Kestin et al \cite{Noam_Kestin_2019}. The thermal ensemble is a mixed state and the matrix product density operator (MPDO) ansatz is better suited for simulating mixed states. However, it is more convenient to construct the MPDO in terms of  purified matrix product states (MPS). 
 A purified MPS is defined on an enlarged Hilbert space, namely a physical Hilbert space and an ancillary Hilbert space \cite{Verstraete_Cirac_2004,Schollwock_2011,Paeckel_2019}. We simulate purified states up to half the inverse temperature denoted as $\hat{\rho}_{\beta/2}$ by applying the time evolution operator in second-order Suzuki-Trotter steps \cite{Feiguin_White_2004}  and then compute the ensemble average of observables by tracing over the ancillary degrees of freedom. Since the thermal ensemble is Hermitian (i.e. $\hat{\rho}_{\beta/2} = \hat{\rho}_{\beta/2}^{\dagger}$), one can write the following expression for computing observables $\mathrm{Tr}\left( \hat{\rho}^{\dagger}_{\frac{\beta}{2}}O\hat{\rho}_{\frac{\beta}{2}}\right)$. Only tracing over ancillary degrees of freedom, i.e. $\mathrm{Tr}_a\left(\hat{\rho}_{\frac{\beta}{2}}\hat{\rho}^{\dagger}_{\frac{\beta}{2}}\right)$, results in the full thermal ensemble ($\hat{\rho}_{\beta}$) as a MPDO.  

The computation of the thermal DSF essentially consists of two steps: (i) simulation of the thermal ensemble by performing imaginary time evolution; (ii) simulation of real time evolution after applying the relevant spin operator to the thermal ensemble. For the simulation of the thermal ensemble, we used imaginary time Trotter steps of  $\Delta \beta = 0.01/J$ keeping the truncation weights to be of the order of $\mathcal{O}\left(10^{-8}\right)$. For real-time evolution we used the trotter steps to be $\Delta t =0.1/J$ keeping the truncation weights to be $\mathcal{O}\left(10^{-4}\right)$. In practice, the time-dependent correlation functions is computed as follows: 
\begin{eqnarray}
C^{\alpha\tilde{\alpha}}(l,t;\beta) = \mathrm{Tr}\lbrack \hat{\rho}^{\dagger}_{\beta/2}S^{\alpha}(\Delta r_l;0)e^{-iHt} S^{\tilde{\alpha}}(0;0)\hat{\rho}_{\beta/2}e^{iHt}\rbrack\nonumber
\end{eqnarray}
Since the space Fourier-transformed correlations for \textit{positive} and \textit{negative} times are related by conjugation, only the positive time correlations were simulated. The simulations were run up to a final time $t_f = 20/J$ (instead of infinite time) to obtain the DSF, which is large compared to the interaction strength ($1/J$). We made sure that, for all the system sizes, the temporal spread of correlations does not reach the boundary. Upon taking the time Fourier transform, the DSF gets convoluted with a sharp window of finite time which results in numerical artefacts. This problem is overcome by multiplying the correlations with a Gaussian filter $2(\pi t^2_f)^{-1/2}e^{-4t^2/t_f^2}$ in order to smooth out the finite time effects \cite{Feiguin_White_2004, Bouillot_2011}. The resulting numerical DSF has a spatial resolution of $\Delta k = 2\pi/L$ and a frequency resolution of $\Delta \omega = \pi/t_f$.

Numerically, the spin-1 chain poses a separate challenge. Since the physical dimension increases and the complexity of the code scales as $\mathcal{O}(d^6\chi^3)$, it is expensive in computational resources. We restricted ourselves to 60 sites and $\chi = 400$ for the thermal DSF computations. For larger number of sites, one would have to keep a larger bond-dimension in order to faithfully represent the thermal ensemble. In order to avoid the boundary effects from the time-evolution cone touching the sides of the chain, we limit ourselves to final time $t_f = 14/J$. This reduces our frequency and spatial resolution as compared to the spin-$1/2$ chain. Since we work with a small chain size and a small final time evolution, we applied a Gaussian filter $2(\pi t^2_f)^{-1/2}e^{-4t^2/t_f^2}e^{-4x^2/(L-1)}$ to the DSF in order to smoothen out both temporal and spatial finite size effects.

\section{Spin-1/2 Ferromagnetic chain}
\label{spin-1/2 FM chain}

\subsection{Model, scattering states, and bound state}

We discuss the spin-1/2 ferromagnetic Heisenberg chain with $L$ sites and periodic boundary conditions described by the Hamiltonian:
\begin{eqnarray}
\mathcal{H}_{FM} = -J\sum_{i=1}^{L} \mathbf{S}_{i}\cdot \mathbf{S}_{i+1}, \hspace{0.4cm}J>0
\label{spin_half_FM_Hamiltonian}
\end{eqnarray}
The low-lying excitations can be described in terms of spin-deviations. 
\begin{enumerate}
\item The ground state can be chosen to be the state which is fully-aligned in a given direction. It has no spin-deviation. The ground state energy is $E_0 = -JL/4 $.  
\item There are $L$ one-spin-deviation states. The Hamiltonian is trivially diagonalized in the momentum basis, leading to spin-wave states with energy:
\begin{eqnarray}
\omega_{1}(k) = E_{1}(k)-E_0 = J(1-\cos k)
\label{spin12_spinwave}
\end{eqnarray}
\item There are $L(L-1)/2$ two-spin deviation states which can be written in the basis of the centre of mass momenta. One can classify them into two kinds of states:
(i) $L(L-3)/2$ scattering states of pairs of spin-waves, and (ii) $L$ bound states of spin-waves. The energy of the scattering states is given by the sum of the energies of two spin-waves:  
\begin{eqnarray}
\omega_{2}(K,p) &=& J(1-\cos k_1)+J(1-\cos k_2)\nonumber\\
&=&2J\left(1-\cos \frac{K}{2}\cos p\right)
\label{spin12_two_spinwave}
\end{eqnarray}
The last line is obtained by transforming the momenta coordinates to the centre of mass momentum $K \equiv k_1+k_2$ and the relative momentum $p \equiv (k_1-k_2)/2$. The bound-state energy can be derived by following the Green's function approach \cite{Wortis_1963} or simply solving the eigenvalue equation \cite{Fukuda_Wortis_1963, DCMattis} (see Appendix \ref{Appendix_spin_half_FM_chain} for details). The dispersion relation of the bound state is: 
\begin{eqnarray}
\omega_{2,\mathrm{BS}}(K) = E_{2,\mathrm{BS}}(K)-E_0 = J \sin^{2}\frac{K}{2}
\label{spin12_boundstate}
\end{eqnarray}
\end{enumerate}
 The difference between the lower boundary of the continuum (given by $p=0$ in Eq.\ref{spin12_two_spinwave}) and the  bound state energy is given by: 
$$
4J\sin^{2}\frac{K}{4} - J\sin^{2}\frac{K}{2} = 4J\sin^{4}\frac{K}{4}\geq 0,
$$ 
with the equality holding only for $K=0$. Thus, the bound state exists as a well separated excitation for all $K>0$.

\subsection{Thermodynamics}

In order to benchmark the first step of the computation of the thermal DSF, we simulate the thermodynamics of the spin-$1/2$ FM chain. Bloch discussed the low-temperature thermodynamics of 3D ferromagnets in terms of non-interacting spin-waves \cite{Bloch}. Since the density of spin-waves in the model is very small at low-temperatures, the spin-waves can be assumed to be non-interacting. This leads to the well known temperature dependence of the magnetisation $M(T)\sim M(0)\left(1-{(T/T_c)}^{\frac{3}{2}}\right)$ for 3D. If one tries to extend this argument to 1D, one gets a diverging correction to the magnetisation. However, from Mermin-Wagner-Hohenberg theorem, the magnetisation of the 1D ferromagnet should be zero. Therefore, Takahashi complemented the non-interacting spin-wave theory with a constraint of zero magnetisation \cite{Takahashi_1986}, a method known as modified spin-wave theory, to explain the low temperature thermodynamics of the 1D ferromagnet. This leads to the following low-temperature expansion of the free energy of the 1D ferromagnet: 
\begin{eqnarray}
F &=& \frac{E_0}{L}-\frac{\zeta\left(\frac{3}{2}\right)}{\sqrt{2\pi}}T^{\frac{3}{2}}+T^2\nonumber\\
&+&\sqrt{\frac{2}{\pi}}\left(\zeta\left(\frac{1}{2}\right) - \frac{\zeta\left(\frac{5}{2}\right)}{16}\right)T^{\frac{5}{2}}+\mathcal{O}(T^{3})
\label{free_energy_density}
\end{eqnarray}
where, $\zeta(\alpha)$ is the Riemann-zeta function. Other interesting thermodynamic quantities such as the entropy, the average energy and the specific heat can be extracted by using statistical physics relations, leading to the low temperature behaviours
\begin{eqnarray}
&&S\propto T^{\frac{1}{2}}\nonumber\\
&&\langle E\rangle - E_0\propto T^{\frac{3}{2}}\nonumber\\
&&C_v \propto T^{\frac{1}{2}}\nonumber
\end{eqnarray} 
Note that the free energy of the 1D ferromagnet has also been computed using thermal Bethe-Ansatz by Takahashi \cite{Takahashi_1971}. This method leads to a set of coupled integral equations on spin-deviations which is analytically solvable in the limit of very low temperature. The agreement between the modified spin-wave theory and the thermal Bethe Ansatz results is excellent at very low temperature \cite{Takahashi_1971,Takahashi_1986}. 

\begin{figure}
\centering
\includegraphics[width = 9cm, height= 9cm ,keepaspectratio]{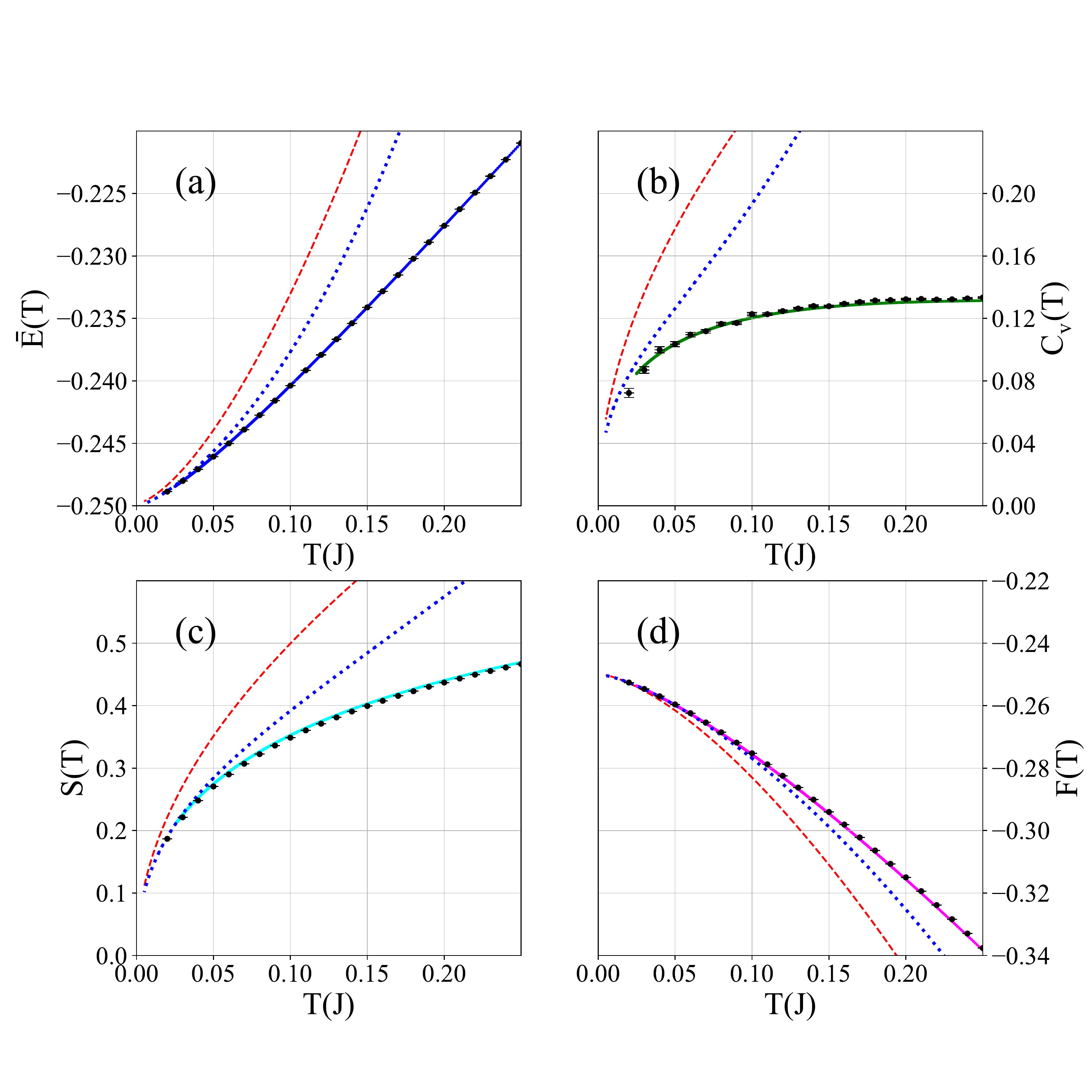}
\caption{Thermodynamics of the spin-1/2 FM Heisenberg chain. The solid lines stand for our thermal DMRG data. The non-interacting spin-wave thermodynamic quantities are shown as red dashed line while the modified spin-wave thermodynamic quantities (low temperature expansion in Eq. \ref{free_energy_density}) are shown as blue dotted lines. The black symbols are QMC data obtained with the Wang-Landau algorithm. See main text for details.}
\label{Thermal_statics_FM}
\end{figure}

Numerically one is limited to finite system sizes of the thermal ensemble and the thermodynamic limit of the thermal quantities is obtained by a finite-size scaling analysis.  The thermal ensemble energy is readily computed from the thermal ensemble and the specific heat is obtained by differentiating the thermal ensemble energy with respect to temperature (see Appendix \ref{finite_size_scaling_therm_quant} for finite-size scaling data). The entropy per unit length is obtained by numerically integrating the specific heat with respect to the inverse temperature. The results for the energy and the entropy lead to the estimation of the free energy. This sequence of steps is summarized below: 
\begin{eqnarray}
&\mathrm{(i)}&\hspace{0.2cm}\langle E\rangle_{\beta} = \frac{1}{L\mathcal{Z}}\mathrm{Tr}\left\lbrack \mathcal{H}e^{-\beta\mathcal{H}}\right\rbrack = \frac{1}{L}\mathrm{Tr}\lbrack\hat{\rho}_{\beta/2}\mathcal{H}\hat{\rho}_{\beta/2}\rbrack\nonumber\\
&\mathrm{(ii)}&\hspace{0.2cm}C_v(\beta) = \frac{d}{dT}\langle E\rangle_{\beta}\nonumber\\
&\mathrm{(iii)}&\hspace{0.2cm}S(\beta) = k_B\int^{\beta}_{\infty}\frac{C_v(\beta)}{\beta}d\beta\nonumber\\
&\mathrm{(iv)}&\hspace{0.2cm}F(\beta) = \langle E\rangle_{\beta}- TS(\beta)\nonumber
\end{eqnarray}
The agreement between our numerics and modified spin-wave theory is very good at low temperatures, where non-interacting spin-waves dictate the thermodynamics (see Fig. \ref{Thermal_statics_FM}). For higher temperatures, interactions between spin-waves become more important and a better agreement with the numerics would probably be obtained using thermal Bethe-Ansatz, but this is beyond the scope of this article. To benchmark our results at not so low temperatures, we have used the Wang-Landau Stochastic Series Expansion QMC code of the ALPS package to calculate these thermodynamic quantities \cite{Wang_Landau_2001,ALPS_Troyer_2003, Bauer_2011} using $L=140$ sites, a cut-off $\Lambda=10^4$, and a temperature step $\Delta T = 10^{-3}(1/J)$. The agreement with our determination of the thermodynamic quantities is perfect within the error bars of the QMC data. 

\subsection{Finite-temperature dynamics}
\begin{figure}
\centering
\includegraphics[width= 14 cm, height = 14 cm, keepaspectratio]{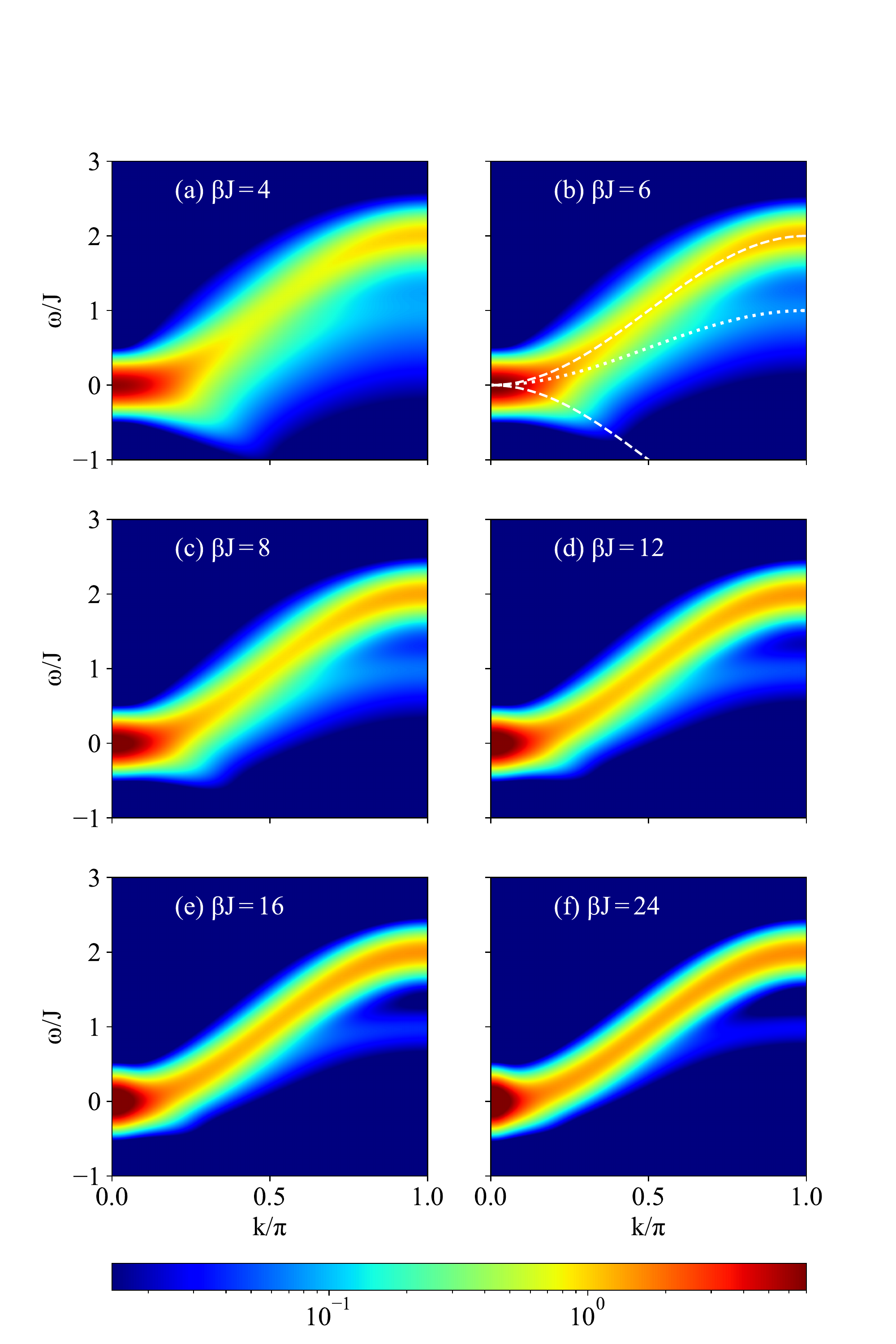}
\caption{Longitudinal thermal DSF (${S^{zz}(k,\omega)}_{\beta}$) of the spin-1/2 FM chain with $L = 140$ sites, $\chi=400$ and $t_f = 20/J$. Since the system is isotropic, the spin-flip DSF component ($S^{xx}(k,\omega)_{\beta}$) is equal to the longitudinal DSF component. As the temperature is lowered (or the inverse temperature $\beta$ increased), the bound state progressively loses spectral weight. The features in Fig. \ref{Finite_temp_without_h_DSF_spin_half}b) are consistent with the spin-wave and bound state dispersion relations of Eq.\ref{spin12_spinwave} and Eq.\ref{spin12_boundstate} which are shown as white dashed and dotted lines respectively. The de-excitation processes from spin-wave state to FM fully aligned state is also denoted in the figure by white dashed line.}
\label{Finite_temp_without_h_DSF_spin_half}
\end{figure}

 From the real-time evolution of the thermal ensemble, we computed the longitudinal component of DSF, shown in Fig. \ref{Finite_temp_without_h_DSF_spin_half}. Since the ferromagnetic chain is isotropic, the transverse components are the same up to a multiplicative factor. At non-zero temperatures, the FM chain develops a thermal population of single-spin-deviation states. Flipping a single spin can result in either of two possibilities: (i) de-excite the state to a fully-aligned state; (ii) create an additional spin-deviation resulting in a two-spin-wave state or a bound state. In the first case, it results in non-zero thermal spectral weight in the negative $\omega$ regime (see Fig. \ref{Finite_temp_without_h_DSF_spin_half}b). The spectral weight for this process is proportional to $e^{-\beta \omega_1(k)}\approx e^{-\beta Jk^2}$ as is evident from Eq. \ref{DSF_formula}. Because of the exponential decay with respect to the wave vector k and inverse temperature, spectral weights are only visible at lower temperatures and close to k = 0. In the second case, since the single spin-flip results in an additional spin-deviation, the bound state can gather significant enough spectral weights to be detected with neutron scattering experiments above some temperature. However, at higher temperatures, the thermal broadening of spin-wave excitations assisted by two-magnon processes obscures the bound state. Therefore, in the thermal DSF, the bound state can only be detected as a separate mode in a suitable range of temperatures near $k = \pi$.\par 

\begin{figure}
\centering
\includegraphics[width =14cm, height = 14cm, keepaspectratio]{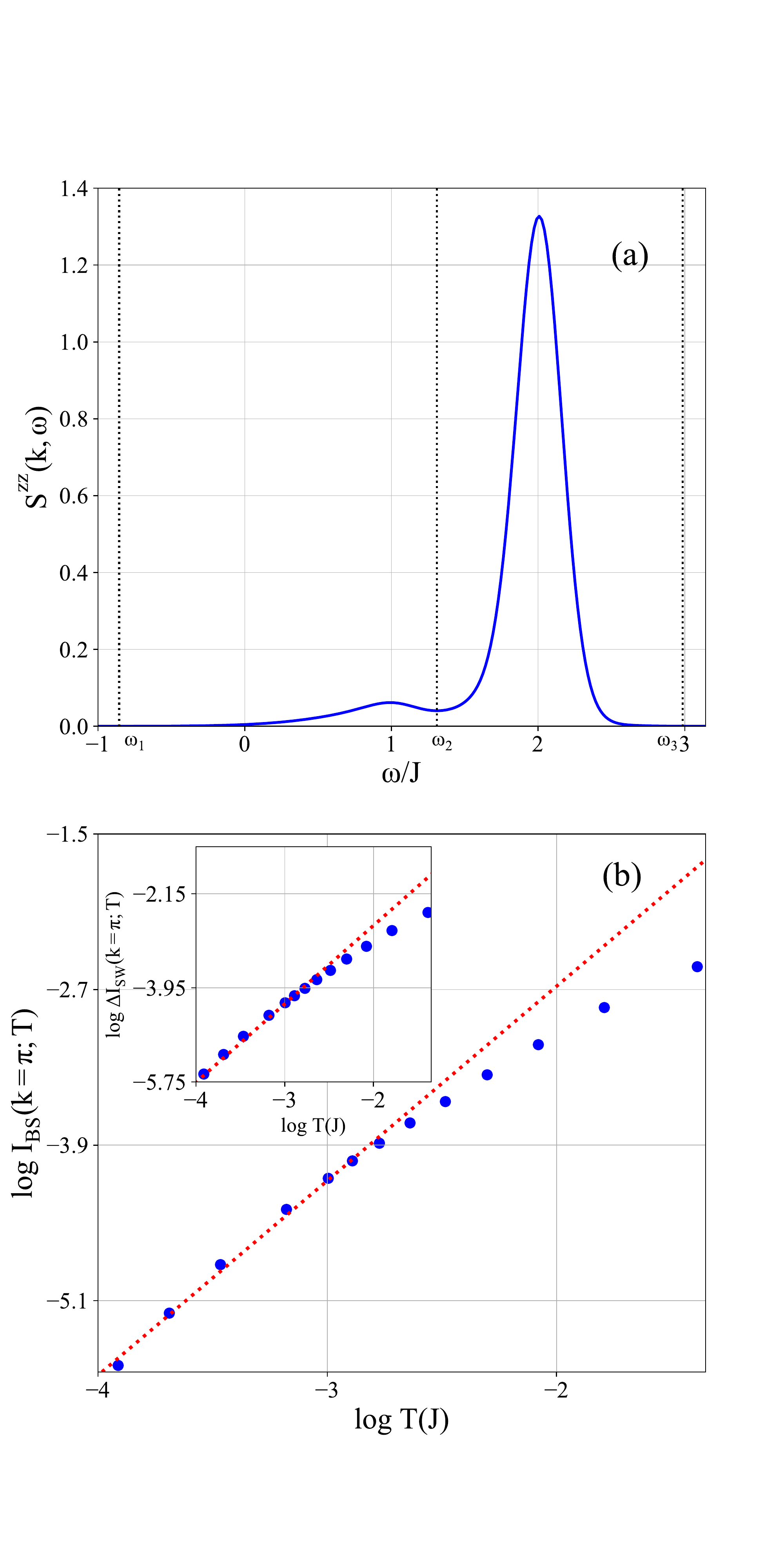}
\caption{(a) Numerical longitudinal thermal DSF section cut of the spin-1/2 FM chain at $k=\pi$ and $\beta J=8$. The spin-wave excitation is the main spectral peak at $\omega = 2J$ and the bound state is the smaller spectral peak at $\omega = J$; (b) log-log plot of the area under the bound state versus temperature (see main text for details). At low temperatures, it is consistent with an exponent 3/2  (red dotted line). Inset: log-log plot of the difference between the area under the main spin-wave excitation at zero temperature and finite temperature versus temperature. At low temperature, it is also consistent with an exponent 3/2  (red dotted line).}
\label{section_cut_area_comparison}
\end{figure}

 To be more quantitative, we considered section cuts for various temperatures at $k =\pi$ (See Fig.\ref{section_cut_area_comparison}). The area under the curve for the bound state ($\mathrm{I}_{\mathrm{BS}}$) and for spin-wave excitations ($\mathrm{I}_{\mathrm{SW}}$) is given by integrating the longitudinal DSF over different frequency ranges (Fig.\ref{section_cut_area_comparison}a):
\begin{eqnarray}
\mathrm{I}_{\mathrm{BS}}(k=\pi;\beta) &=& \int_{\omega_1}^{\omega_2}S^{zz}(k=\pi,\omega)_{\beta} d\omega\nonumber \\
\\
\mathrm{I}_{\mathrm{SW}}(k=\pi;\beta) &=& \int_{\omega_2}^{\omega_3}S^{zz}(k=\pi,\omega)_{\beta} d\omega,\nonumber 
\end{eqnarray}
where $\omega_1$ and $\omega_3$ are such that the section-cut curve lies below $10^{-4}$ outside this range, while $\omega_2$ is the value where the section cut curve reaches a local minimum between the bound state peak and the main spin-wave excitation peak.  The fraction of total spectral weight under the bound state in the section cut forms a direct criterion for it to be detectable in the INS experiment. The feature is considered to be visible if it gathers more than 5 percent of the total spectral weight at $k = \pi$, which sets the lower limit of the temperature range. For the upper limit, we use the criterion that the thermal broadening of the bound state and of the main spin-wave mode are such that the bound state is no longer distinguishable. Therefore, from these criteria, the bound state feature can be detected in the temperature regime $J/12<k_BT<J/3$ (Appendix \ref{temp_range}).\par
Since the DSF is simulated for finite sizes, the thermal spectral areas at $k=\pi$ have to be extrapolated in $1/L$ to obtain the spectral areas in the thermodynamic limit (Appendix \ref{finite_size_scaling_bound_state}). A log-log plot of these extrapolated areas is shown in Fig. \ref{section_cut_area_comparison}b.  It is difficult to push the thermal DMRG algorithm to the $T=0$ limit since one would have to simulate upto $\beta J=\infty$. However, for $T=0$, the DSF should capture only the spin-wave excitation, so the area under the bound state curve is $\mathrm{I}_{\mathrm{BS}}(k=\pi;\beta=\infty) = 0 $. We determined the area under the main spin-wave excitation at $T=0$ by ensuring that the sum of the area of the spin-wave peak and of the bound state peak is constant. We make the following observations at this section-cut :\\
 (a) The spectral peak of the bound state has a longer tail that extends to low energies (see section \ref{TDSF_extr_mag_field}, \ref{Section_cut_simple_calculation}).\\
 (b) The area under the spectral peak associated with the bound state scales  with temperature as $T^{\frac{3}{2}}$ (Fig. \ref{section_cut_area_comparison}b). This is because the thermal ensemble population of spin-waves dominantly scale as $T^{\frac{3}{2}}$. It also indicates that the spectral weight of spin-waves arising from fully aligned states would be decreasing in a similar way. We verified this from plotting the following quantity (inset of Fig. \ref{section_cut_area_comparison}b):
\begin{equation}\Delta I_{\mathrm{SW}}(k=\pi;\beta) = I_{\mathrm{SW}}(k=\pi; \infty) -I_{\mathrm{SW}}(k=\pi;\beta)\nonumber\end{equation}
These observations can be qualitatively supported by simple calculations in the presence of a vanishing magnetic field  (see section \ref{Thermal_DSF_mag_field}).

\begin{figure}
\centering
\includegraphics[width = 9.25cm ,height = 9.25cm ,keepaspectratio]{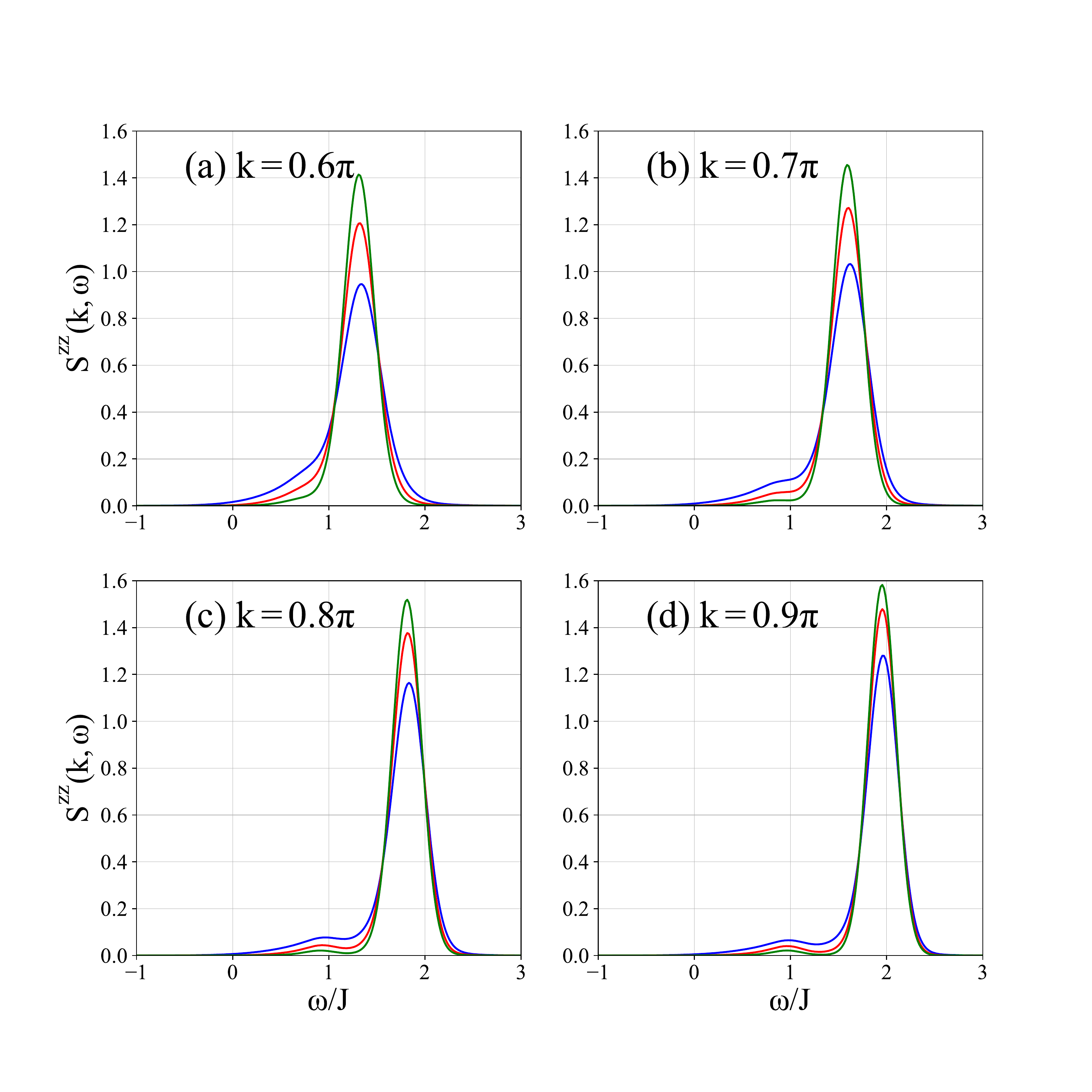}
\caption{Comparison of section cuts of the longitudinal thermal DSF of the spin-1/2 FM chain with 140 sites for $\beta J  = 8$ (blue), $\beta J =16$ (red) and $\beta J = 32$ (green). As the temperature decreases, the spectral peak associated with the bound state decreases in height and gets separated from the single spin-wave excitation, but for section-cuts at $k = 0.6\pi$ (shown in a) and $k=0.7\pi$ (shown in b), the bound state is not completely separated even at very low temperatures. }
\label{Comparison_section_cuts_LDSF}
\end{figure}

Finally, in Fig. \ref{Comparison_section_cuts_LDSF}, we plot the section-cuts of the longitudinal component of the thermal DSF at various wave-vectors. As the wave-vector increases towards $\pi$, the bound state mode is completely separated for low enough temperatures. It is interesting to note that for $k =0.6\pi$ and $k = 0.7\pi$, the bound state is not completely separated from the two-magnon continuum. The presence of the bound state appears as an asymmetric spectral peak with a long tail.
\section{ Thermal DSF in a magnetic field}
\label{Thermal_DSF_mag_field}
We study the problem in the presence of an external magnetic field defined by the Hamiltonian: 
\begin{eqnarray}
H_{FM,h} = - J \sum_{i} \mathbf{S}_{i}\cdot \mathbf{S}_{i+1} +h\sum_{i}S^{z}_{i}
\end{eqnarray}
Since single-spin-deviation states gain Zeeman energy  $h$ and two-spin-deviation states gain Zeeman energy  $2h$, the dispersion relation for spin-wave and the dispersion relation for the bound state becomes
\begin{equation}
\omega_{1,h}(k) = J(1-\cos k)+h
\label{disperson_reln_mag_field_spin_wave}
\end{equation}
\begin{equation}
\omega_{2,BS,h}(k) = J\sin^2\frac{k}{2}+2h
\label{disperson_reln_mag_field_bound_state}
\end{equation}
The model is no longer isotropic, therefore, it has 3 different thermal DSF components namely - (i) Longitudinal component $S^{z,z} (k,\omega)_\beta$, (ii) Transverse component $S^{+,-}(k,\omega)_\beta$ and (iii) Transverse component $S^{-,+}(k,\omega)_\beta$. 

\subsection{Longitudinal Dynamical Structure Factor in a magnetic field}
\label{LDSF_extr_mag_field}
\begin{figure*}
\centering
\includegraphics[width =16cm, height =16cm, keepaspectratio]{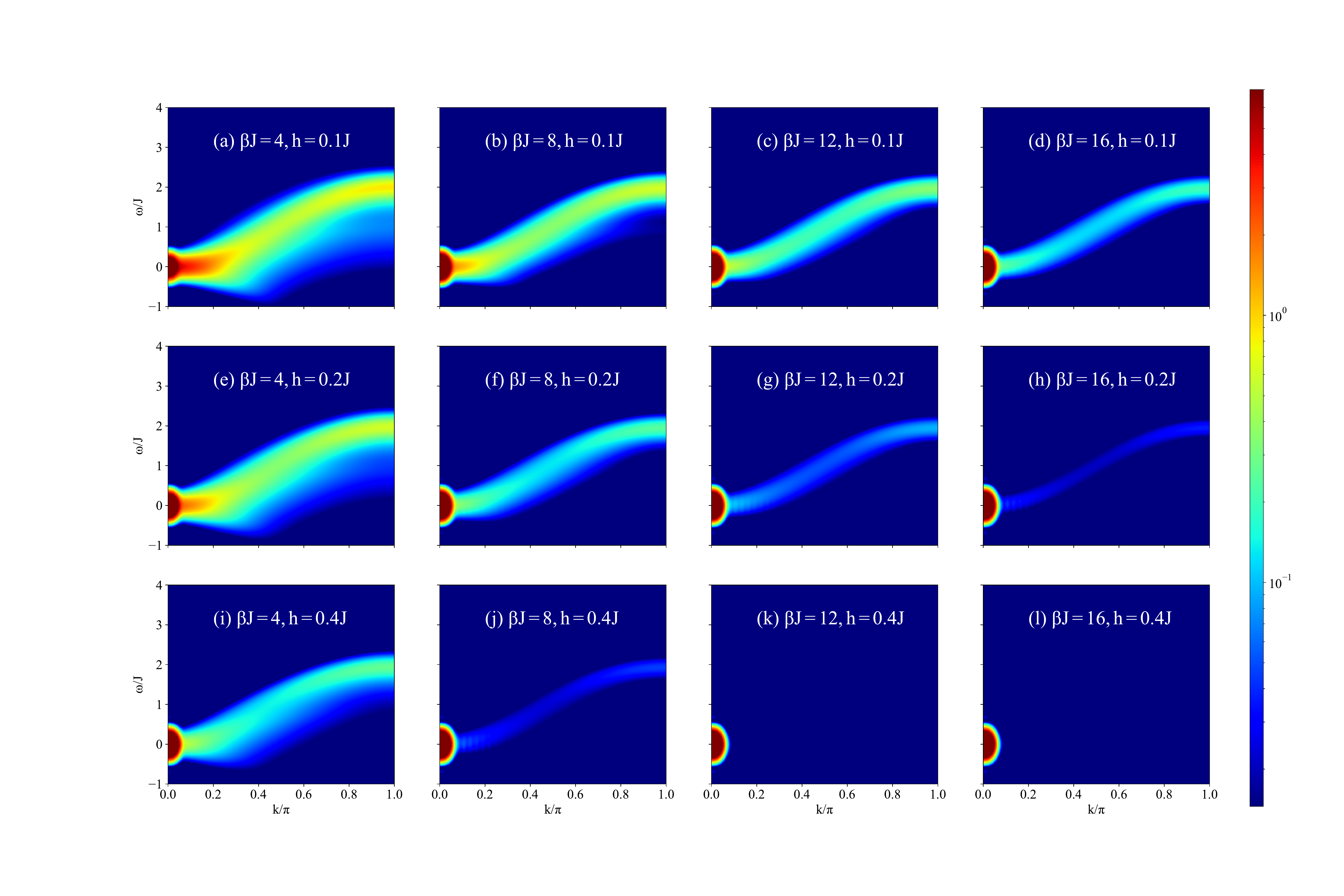}
\caption{Longitudinal thermal DSF($S^{zz}(k,\omega;\beta)$) of the spin-1/2 FM chain in the presence of a magnetic field ($L =140$, $t_f = 20/J$). The spectral weight gathered by the spin-wave dispersion is one-tenth of the spectral weight at $\omega =0 ,k=0$. Signatures of the bound state are visible, but they are one-hundredth the spectral weight of the main feature. }
\label{Longitudinal_DSF_h}
\end{figure*}

 The degeneracy of the FM ground state is lifted upon applying an external magnetic field and the ground state of Hamiltonian $\mathcal{H}_{FM,h}$ is given by the fully polarised state in the direction opposite to the magnetic field. The zero-temperature longitudinal component of the DSF is obtained by determining the time-dependent correlation between the $z$-components (denoted by $C_{zz}(m,n;t)$) and then taking both space and time Fourier transform
\begin{eqnarray}
C_{zz}(m,n;t) &=& \bra{\mathrm{GS}}S^{z}_{m}(t)S^z_{n}\ket{\mathrm{GS}}\nonumber\\
S^{zz}(k,\omega)&=&\frac{1}{L^2}\int_{-\infty}^{\infty} dt\sum_{m,n}C_{zz}(m,n;t)e^{-ik(r_n-r_m)}e^{-i\omega t}\nonumber
\end{eqnarray}
As the groundstate is a fully polarised state in the negative direction, the longitudinal component of the DSF can be directly evaluated to be:
\begin{eqnarray}
S^{zz}(k,\omega) = \frac{\pi}{2}\delta(\omega)\delta(k)
\end{eqnarray}
 As a result, at zero temperature, all the spectral weight is concentrated at $\omega =0, k = 0$. Upon increasing the temperature, the spin-wave state gathers spectral weight with more weight still concentrated near $\omega, k\approx 0$. This is also true if the magnetic field were decreased for a given temperature. Since the weights near $\omega=k=0$ are at least one order of magnitude more than the rest of the $\omega$ or $k$ values, it is very unlikely to observe the spin-waves or bound state at finite temperature INS experiments in the longitudinal channel. We summarise our findings on the longitudinal component of the DSF in Fig. \ref{Longitudinal_DSF_h}.

\subsection{Transverse Dynamical Structure Factor in a magnetic field}
\label{TDSF_extr_mag_field}
\begin{figure*}
\centering
\includegraphics[width =16cm, height =16cm, keepaspectratio]{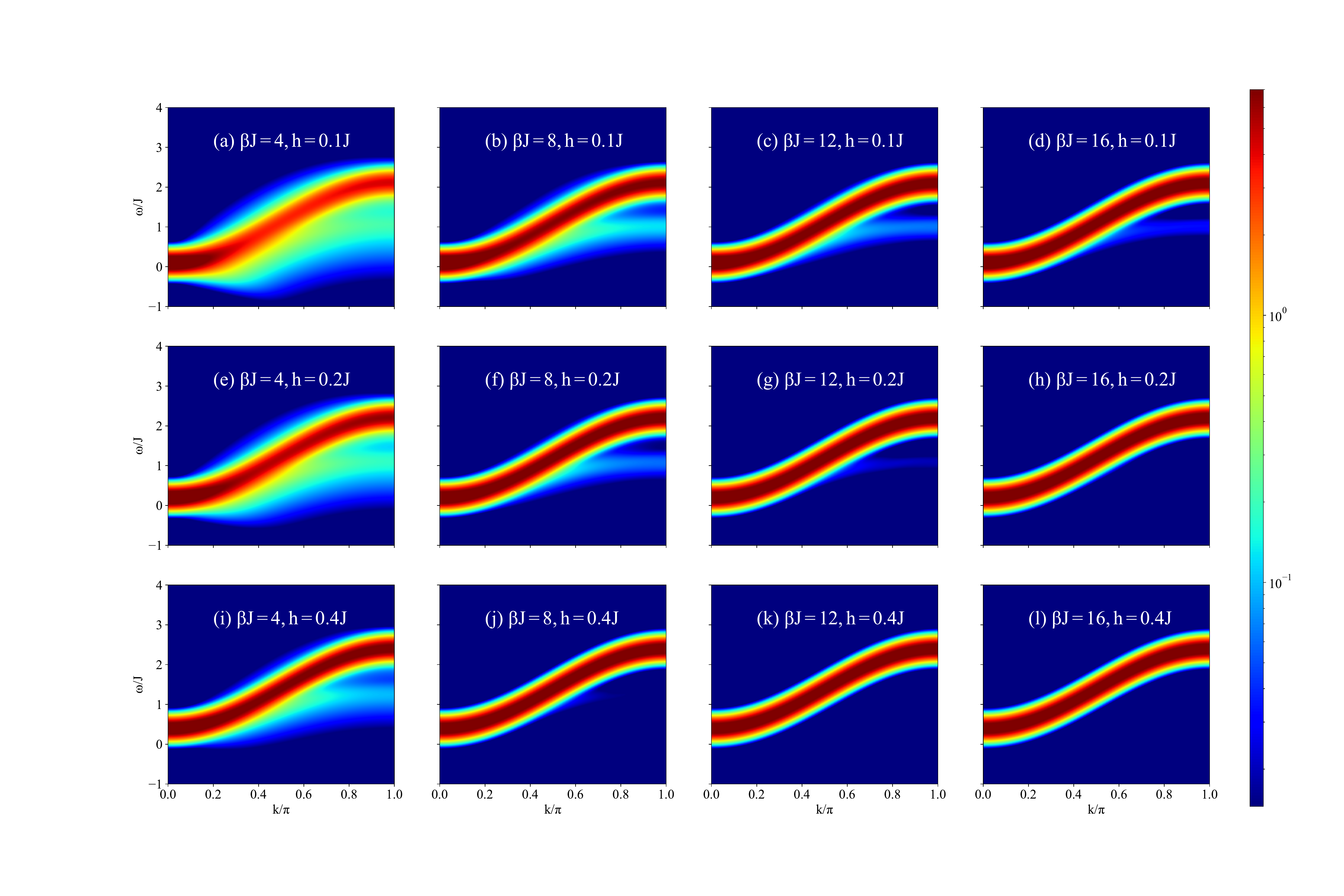}
\caption{Transverse thermal DSF ($S^{-,+}(k,\omega;\beta)$) of the spin-1/2 FM chain in the presence of a magnetic field ($L =140$, $t_f = 20/J$). The spin-wave dispersion is shifted in energy by $h$ (the strength of the magnetic field). The bound state at higher temperatures and lower magnetic fields gathers more spectral weights. It must be noted that these spectral weights are much smaller in magnitude than in the case of zero-magnetic field.}
\label{Transverse_DSF_h}
\end{figure*}

We now present the numerical results of the transverse component $S^{-,+}(k,\omega)_\beta$  where a clear signature of bound state is obtained (see Fig.\ref{Transverse_DSF_h}). A simple calculation can be attempted by only keeping the most dominant terms in the thermal DSF at low temperatures - the single excitation on fully-aligned-state (denoted as $|GS\rangle$) leading to a spin-wave state (denoted as $|\gamma_1\rangle$) and the excitation from a single spin-wave state to a bound state of two spin-waves (denoted as $|\alpha_{\mathrm{BS}}\rangle$). The thermal DSF can be expressed in the Lehmann representation as:
\begin{eqnarray}
S^{-,+}(k,\omega)_{\beta} = \frac{2\pi}{L\mathcal{Z}}\sum_{\eta,\gamma} e^{-\beta E_{\gamma}}{\left |\bra{\eta}S^{+}_{-k}\ket{\gamma}\right |}^2 \delta{\left(\omega-\lbrack \omega_{\eta}-\omega_{\gamma}\rbrack\right)}\nonumber\\
\label{DSF_formula}
\end{eqnarray}
where $\mathcal{Z}$ is the partition function and $|\eta\rangle$ or $|\gamma\rangle$ are eigen states of the model.
Thus, the thermal DSF becomes: 
\begin{eqnarray}
&&S^{-,+}(k,\omega)_{\beta} \propto \frac{e^{-\beta E_0}}{\mathcal{Z}}\Bigg(\sum_{\gamma_1}{\left |\bra{\gamma_{1}}S^{+}_{-k}\ket{\mathrm{GS}}\right |}^2\delta(\omega - \omega_{1, h})+\nonumber\\
&&\sum_{\alpha_{\mathrm{BS}};\gamma_{1}}e^{-\beta\omega_{1,h}}{\left |\bra{\alpha_{\mathrm{BS}}}S^{+}_{-k}\ket{\gamma_{1}}\right |}^2\delta(\omega - \lbrack \omega_{2,\mathrm{BS}, h}-\omega_{1, h}\rbrack)+\dots\Bigg)\nonumber\\
\label{DSF_formula_expansion}
\end{eqnarray} 

The first term inside the bracket is easily calculated: 
\begin{eqnarray}
\sum_{\gamma_{1}}{\left |\bra{\gamma_1}S^{+}_{-k}\ket{\mathrm{GS}}\right |}^2\delta(\omega - \omega_{1,h}(p))=\delta(\omega - \omega_{1,h}(k))\nonumber
\end{eqnarray}
As expected the spin-wave excitation is shifted by $h$. It is thermally broadened in the plots (Fig.\ref{Transverse_DSF_h}),
 but the analytical computation does not capture it here.

The second term in the DSF expression indicates that the spectral weights of the bound state observed at a given wave vector $k$ is due to a finite overlap with the bound state of momentum $K=k+p$, where $p$ is the momentum of a thermally excited spin-wave. So, for computing the second term, we replace the sum over bound states and spin-wave states with a sum over $K$ and $p$ respectively. The bound state can be expressed in terms of two spin-deviations (see Eqs. \ref{two_spin_wave_state} and \ref{coeff_bound_state} in Appendix \ref{Appendix_spin_half_FM_chain}). $S^{-,+}(k,\omega)_{\beta,2}$ is thus proportional to:
\begin{eqnarray}
&&\sum_{K,p}e^{-\beta\omega_{1,h}(p)}{\left |\bra{\alpha_{\mathrm{BS}}}S^{+}_{-k}\ket{\gamma_{1}}\right |}^2\delta(\omega - \lbrack \omega_{\mathrm{BS},h}(K)-\omega_{1,h}(p)\rbrack)\nonumber\\
&=&\frac{16}{L}\sum_{p}e^{-\beta\omega_{1,h}(p)}{\left |\sin \frac{p+k}{2}\right |}^2{\left | f_s(p+k;p-k)\right |}^2\nonumber\\
&&\delta(\omega - \lbrack \omega_{\mathrm{BS},h}(p+k)-\omega_{1,h}(p)\rbrack)\nonumber
\end{eqnarray}
where, in the thermodynamic limit  ($L\gg1$), the factor $f_{s}(p+k;p-k)$ is given by:
$$
\frac{\cos \left(\frac{p-k}{2}\right)-\cos\left(\frac{p+k}{2}\right)}{3+\cos \left(p+k\right) -2 \cos p -2 \cos k}
$$  
We selected the section cut at $k =\pi$ to compare our analytical results with the numerical results. In the thermodynamic limit we evaluate the integral in $p$ to compute $S^{-,+}(k=\pi,\omega)_{\beta,2}$, which is proportional to:
\begin{eqnarray}
&&\frac{16e^{-\beta h}}{2\pi}\int_{p=-\pi}^{p=\pi}dpe^{-2\beta J\sin^{2}\frac{p}{2}}\frac{\sin^{2}\frac{p}{2}\cos^{2}\frac{p}{2}}{{\left(1+3\sin^{2}\frac{p}{2}\right)}^2}\times\nonumber\\
&&\delta\left(\omega -J\left(1 - 3\sin^{2}\frac{p}{2}\right)-h\right)
\label{Integral}
\end{eqnarray}
The integral (Eq. \ref{Integral}) is even about $p=0$ and upon doing a change of variable $t = \sin^2\frac{p}{2}$, it becomes: 
\begin{eqnarray}
\frac{16e^{-\beta h}}{3\pi J}\int_{0}^{1}dt e^{-2\beta J t}\left\lbrace\frac{\sqrt{t(1-t)}}{{\left(1+3t\right)}^2}\right\rbrace\delta\left(t-\left(\frac{-\omega + J+h}{3J}\right)\right)\nonumber
\end{eqnarray}
which, for $-2J+h\leq \omega \leq J+h$, leads to:
\begin{equation}
\frac{16e^{-\beta h}}{9\pi}e^{-\frac{2}{3}\beta \left(J+h-\omega\right)}\frac{\sqrt{\left(J+h-\omega\right)\left(2J-h+\omega\right)}}{{\left(\omega-2J-h\right)}^2}
\label{Transverse_analytical_DSF}
\end{equation}
The above expression clearly shows that the spectral weights of the bound state in the section cut extends from the spectral peak to negative energies. Note that there is an exponential suppression of the thermal spectral weights due to the magnetic field. 

\subsection{Section cut in the limit of a vanishing magnetic field}
\label{Section_cut_simple_calculation}
\begin{figure}
\centering
\includegraphics[width = 14cm, height =14cm,keepaspectratio]{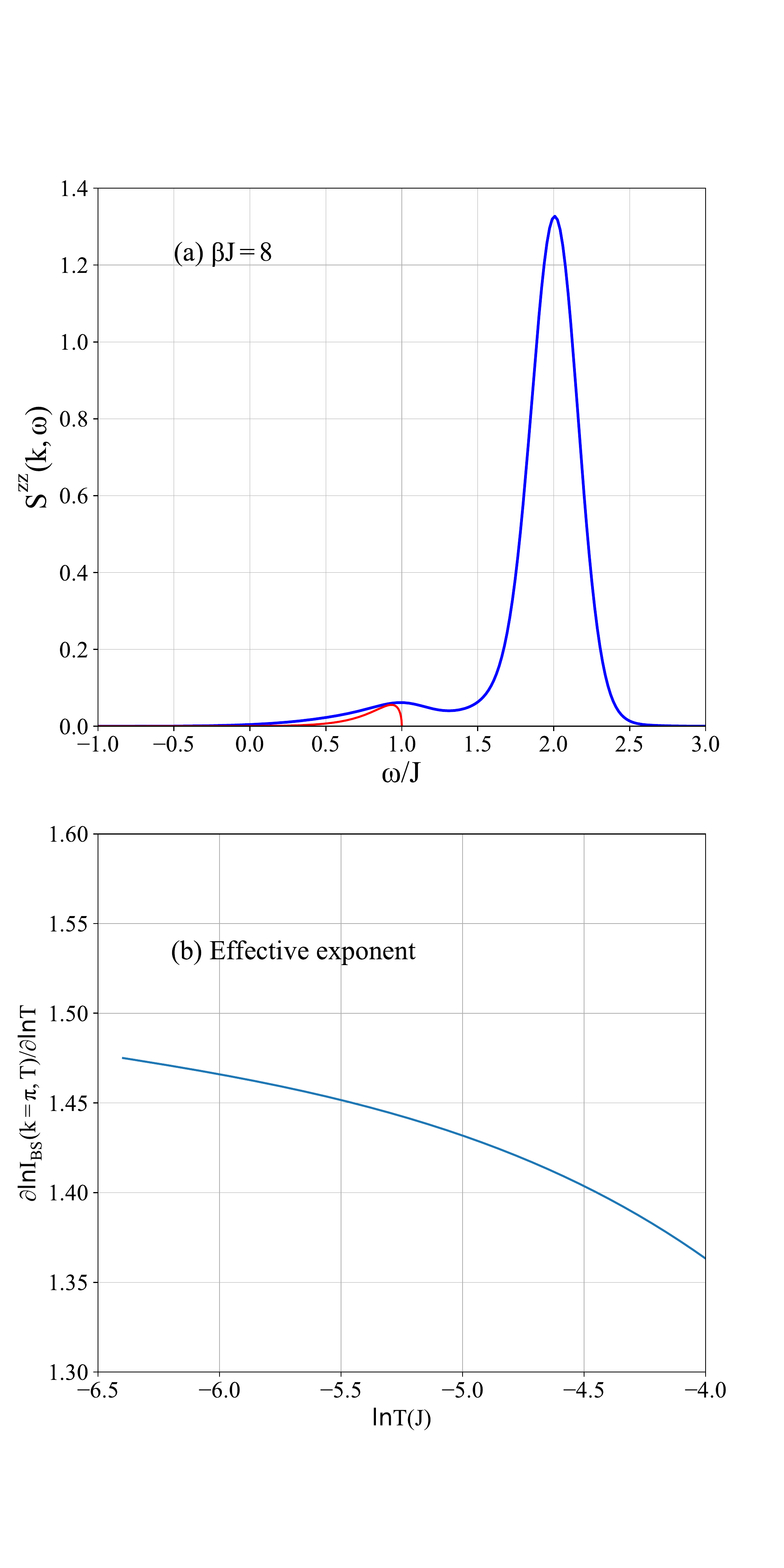}
\caption{(a) Comparison of the numerically determined t-DMRG section cut of the spin-1/2 FM chain at $k=\pi$ (blue) and a simple estimate of the bound state thermal weight (red). The simple estimate misses spectral weight because we only included excitations with a single spin-wave state in the thermal population. 
(b) Effective exponent determined by taking the derivative of the log of the area under the curve with respect to the log of the temperature.}
\label{section_cut_simple_calc}
\end{figure}
In the limit of $h$ going to $0$, the system becomes isotropic and the transverse component is simply equal to two times the longitudinal component of the thermal DSF. One can determine the prefactor (in Eq. \ref{DSF_formula_expansion}) numerically from the free-energy density (in Fig. \ref{Thermal_statics_FM}d) using the formula $F = -k_BT\ln\mathcal{Z}$. 
The expression (Eq.\ref{Transverse_analytical_DSF}), after multiplying by the prefactor, qualitatively explains the long tail of the spectral peak associated with the bound state seen in the numerical thermal DSF section cut (in Fig. \ref{section_cut_simple_calc}a). For very low temperatures, the free-energy density can be replaced with the modified spin-wave theory free-energy density formula (Eq. \ref{free_energy_density}) and the resulting integral can be determined numerically. Its temperature dependence is consistent with $T^{\frac{3}{2}}$ as shown in Fig. \ref{section_cut_simple_calc}b. The effective exponent as a function of temperature is given by the derivative of the log of the area under the bound state with respect to log T [\onlinecite{Timothy}]. At very low temperatures, the exponent tends towards 3/2. Note that this calculation does not include contributions from multi-spin-wave scattering states to the bound state. This is presumably why the effective exponent only tends to the expected value at very low temperatures.

\section{Spin-1 Ferromagnetic chain}
\label{spin-1 FM chain}  
\subsection{Zero temperature dynamics}
The analysis of the spin-1/2 FM chain can be extended to the spin-1 FM chain. A subtlety arises however because two spin-deviations can occupy the same site. To better describe the features in the finite temperature DSF, we will compare it to excitations in the two-spin-deviation subspace. The dispersion relation for a spin-wave excitation in the spin-1 chain is $2J(1-\cos k)$, where $J$ is the interaction strength and $k$ is the momentum. The two-spin-deviation subspace has three types of solutions - (i) two-spin wave scattering states, (ii) bound state and (iii) anti-bound state. 
\begin{enumerate}
\item The energies of the two spin wave scattering states (labelled by $k_1$ and $k_2$)  in the basis of the centre of mass momenta are given by:
\begin{eqnarray}
\omega_2(K,p) = 4J\left(1-\cos\frac{K}{2}\cos p\right),\nonumber
\end{eqnarray}
where $K\equiv k_1+k_2$ and $p\equiv (k_1-k_2)/2$. 
\item The bound state solution can be determined by setting up the transfer matrix equation, as in the spin-$1/2$ case (see appendix A) and by looking for the \textit{localised} solution \cite{Wortis_1963, DCMattis,Tonegawa_1970,Haldane_1982a, Haldane_1982b}. In the thermodynamic limit, the dispersion relation of the bound state in the spin-$1$ case is given by: 
\begin{equation}
\omega_{2,\mathrm{BS}}(K) = \frac{11J}{3}+\frac{J}{3}\left(\frac{13+12\cos K}{\textrm{x}}+\textrm{x}\right)
\end{equation}
with
\begin{eqnarray}
\textrm{x}^3&=&-100-126 \cos K-27\cos 2K\nonumber\\
&+&12\sqrt{6}\sqrt{{\left(\cos\frac{K}{2}\right)}^6(29+27\cos K)} \nonumber
\end{eqnarray}
The difference between the lower energy boundary of the continuum and the bound state energy is positive for $K>0$, so it exists as a well separate excitation. 
\item  While the first two types of excitations are also present in spin-1/2 case, there arises the possibility of an anti-bound state due to two spin-deviations occupying the same site\cite{Papanicolaou_1988}. In the case of the Heisenberg model, this does not give rise to an anti-bound state but to a \textit{resonance} because the energy of that state, which is dispersionless and given by:
\begin{eqnarray}
\omega_{2,\mathrm{ABS}} (K) = 4J
\end{eqnarray}
to first order in the transverse part of the Hamiltonian, overlaps with the two-magnon continuum. If a biquadratic interaction is included however, an anti-bound state well separated from the two-magnon continuum appears (see below).
\end{enumerate}

\begin{figure}
\centering 
\includegraphics[width =7.6cm ,height =15.2cm , keepaspectratio]{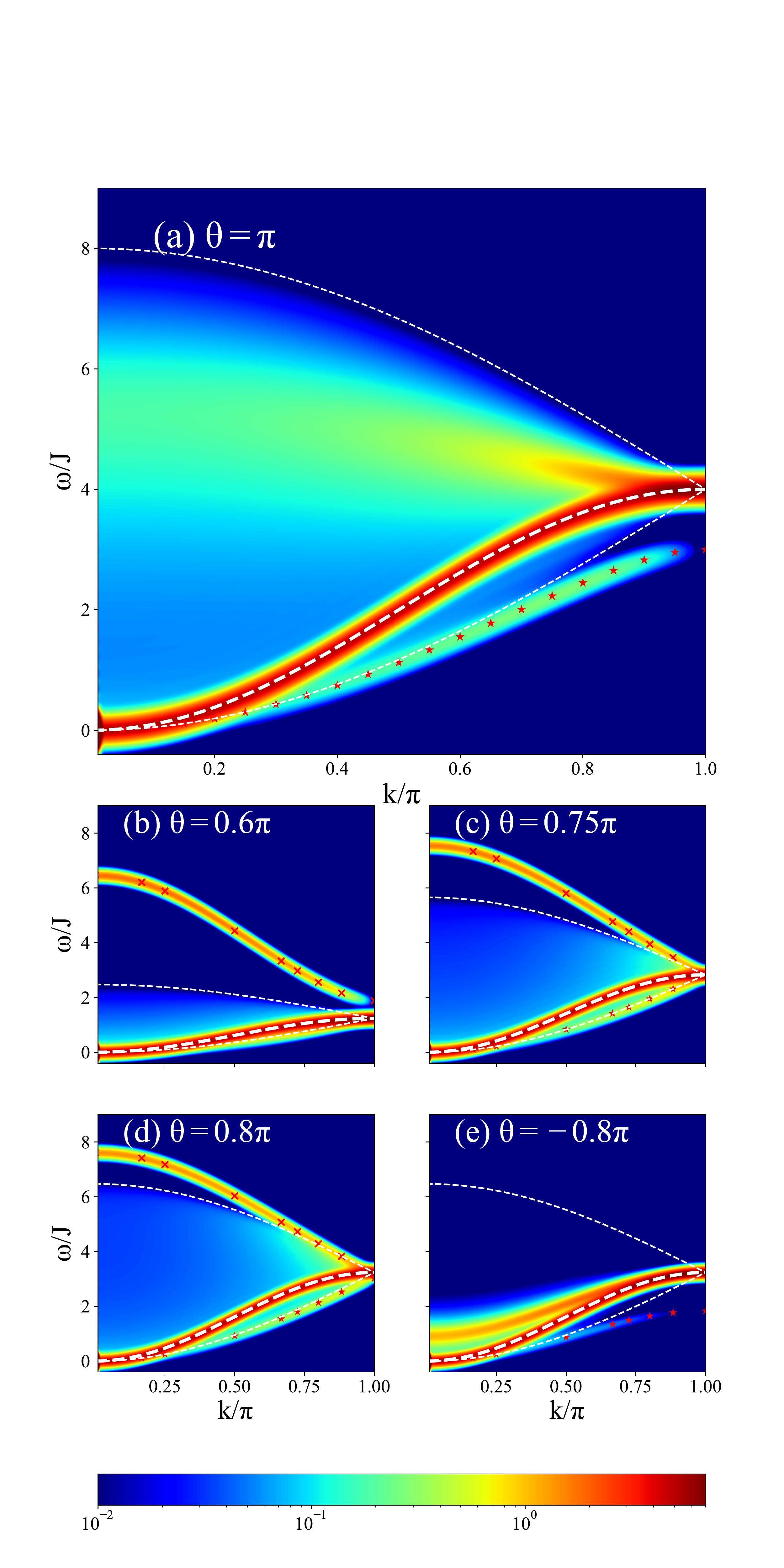}
\caption{ DQSF plots for (a) the spin-1 Heisenberg model, (b-e) the spin-1 BLBQ model ($L =240$, $t_f =24/J$, $T=0$). We find that the bound state (red stars) is more prominent as $k$ goes to $\pi$, but for lower values of $k$ it lies within the two-magnon continuum (bounded by the thin white dashed lines). By contrast, the anti-bound state (red crosses) is more prominent for lower values of $k$ and in some cases merge with the two-magnon continuum for larger values of $k$. The thicker white dashed lines indicate the single magnon dispersion.}
\label{DQSF_BLBQ}
\end{figure}

To numerically characterise the two-spin deviation spectrum, it is useful to look at the zero temperature dynamical quadrupolar structure factor (DQSF). The quadrupolar structure factor has 3 equivalent components for an isotropic system corresponding to the change of on-site magnetisation  ($\Delta S^z$), namely - longitudinal component ($\Delta S^z =0$), transverse component ($\Delta S^z=\pm 1$) and pairing component ($\Delta S^z=\pm 2$) \cite{Salvatore_2011}. We present the numerical results of the pairing component (denoted as $C_{Q,2}(i,j;t)$) but we verified that the three components give the same spectral features up to a multiplicative factor. To be specific, the pairing component of the DQSF is defined by:
\begin{eqnarray}
&&C_{Q,2}(i,j;t) = \frac{1}{2}\left\lbrack\langle{\left(S^{-}(j;t)\right)}^2{\left(S^{+}(i)\right)}^2\rangle+\mathrm{h.c.}\right\rbrack\nonumber\\
&&\\
&&Q_{2}(k,\omega) = \frac{1}{L^2}\int_{-\infty}^{\infty} dt e^{i\omega t}\sum_{i,j}e^{-ik(r_{j}-r_{i})}C_{Q,2}(i,j;t)\nonumber
\end{eqnarray}
where, $Q_{2}(k,\omega)$ is fourier transform of the pairing component. There is a clear evidence of a resonance extending into the two-spin-wave continuum around $k=\pi$ (in Fig. \ref{DQSF_BLBQ}a).

In order to convince ourselves that it is the anti-bound state that results in a resonance, we extend our model by including a nearest neighbour biquadratic coupling. The Hamiltonian is given as:
\begin{equation}
\mathcal{H}_{\mathrm{BLBQ}} = J\sum_{i}\cos\theta\left(\mathbf{S}_{i}\cdot\mathbf{S}_{i+1}\right)+\sin\theta{\left(\mathbf{S}_{i}\cdot\mathbf{S}_{i+1}\right)}^2
\end{equation}
The ferromagnetic phase for this model is located in the interval $\theta\in\left(\pi/2,5\pi/4\right)$. The spin-1 Heisenberg FM chain is recovered for $\theta = \pi$. We simulate the DQSF in the FM phase of the BLBQ model for $\theta\in\left(\pi/2, 3\pi/4\right\rbrack$ and, for $\theta$ not too large, there is clear evidence of an anti-bound state above the two-spin-wave continuum. Following the Green's function approach by Wortis \cite{Wortis_1963}, we found the same dispersion relations for the bound states and anti-bound states as has been obtained for this model by Aghahosseini et. al \cite{Aghahosseini_Parkinson_1978}. They agree with the numerical determination of the spectrum in Fig. \ref{DQSF_BLBQ}. As $\theta$ increases from $\pi/2$ towards the Heisenberg FM point, DQSF plots show that the anti-bound state enters the continuum, explaining the presence of a resonance at the Heisenberg point. Interestingly for negative biquadratic interaction (in Fig. \ref{DQSF_BLBQ}e), there is a stronger decay of spectral weights within the limits of the two-magnon continuum when compared to the positive values of biquadratic interactions. Beyond the two-magnon continuum, the intensity is smaller than the numerical errors (i.e $10^{-5}$). Due to the chosen color scheme, the figure appears to have no weight in the two-magnon continuum. 

\subsection{Finite temperature dynamics}
\begin{figure}
\centering
\includegraphics[width= 9.25 cm , height = 9.25cm, keepaspectratio]{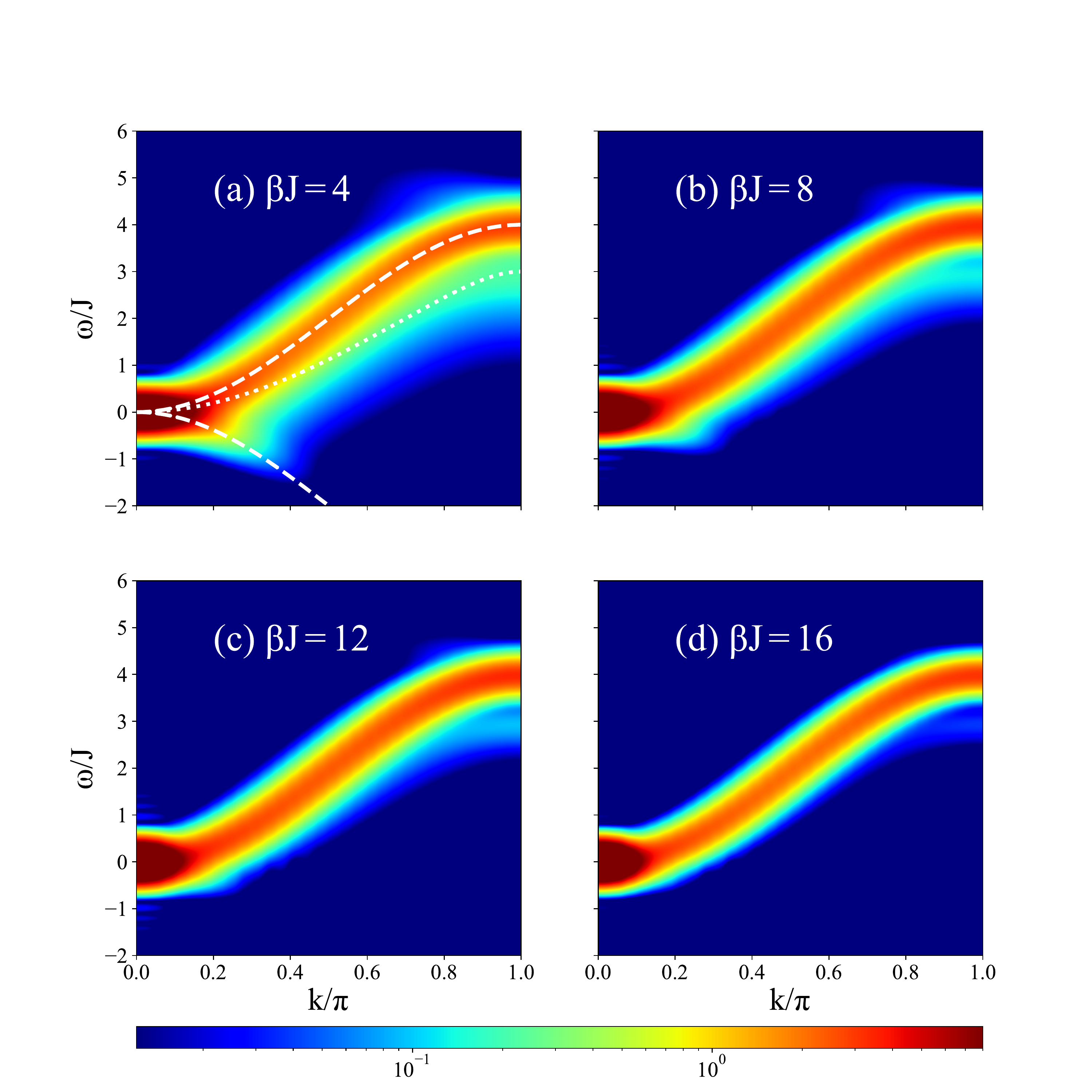}
\caption{Thermal DSF of the spin-1 FM chain at different finite temperatures. The thermal DSF simulation has been carried out with $L = 60$ and $t_f = 14/J$. As compared to the spin-1/2 chain, the resolution of the modes is smaller in the case of the spin-1 chain. In the thermal DSF, the white dashed lines show the spin-wave excitation, while the white dotted line shows the bound state. As in the case of the spin-1/2 chain, the de-excitation process (denoted by the white dashed line) of the spin-wave state to the fully aligned state is also captured here.} 
\label{Spin1_thermal_DSF}
\end{figure}

Similarly to spin-1/2 FM chain, one can expect the bound state in a spin-1 FM chain to be experimentally detectable at finite temperature. And indeed, the numerical simulations of the thermal DSF clearly show the presence of a bound state (see Fig. \ref{Spin1_thermal_DSF}). The energy difference between the single spin-wave excitation and the bound state of two spin-waves is maximum at $k =\pi$, like in the spin-1/2 chain, and it is equal to $J$, so it should be possible to resolve it. 
In the infrared-limit, even though the bound state only touches the spin-wave continuum at $k =0$, the bound state feature is cloaked because of the thermal broadening of the single spin-wave excitation and of two-spin-wave processes. \par
\begin{figure}
\centering
\includegraphics[width =9.25cm, height =9.25cm, keepaspectratio]{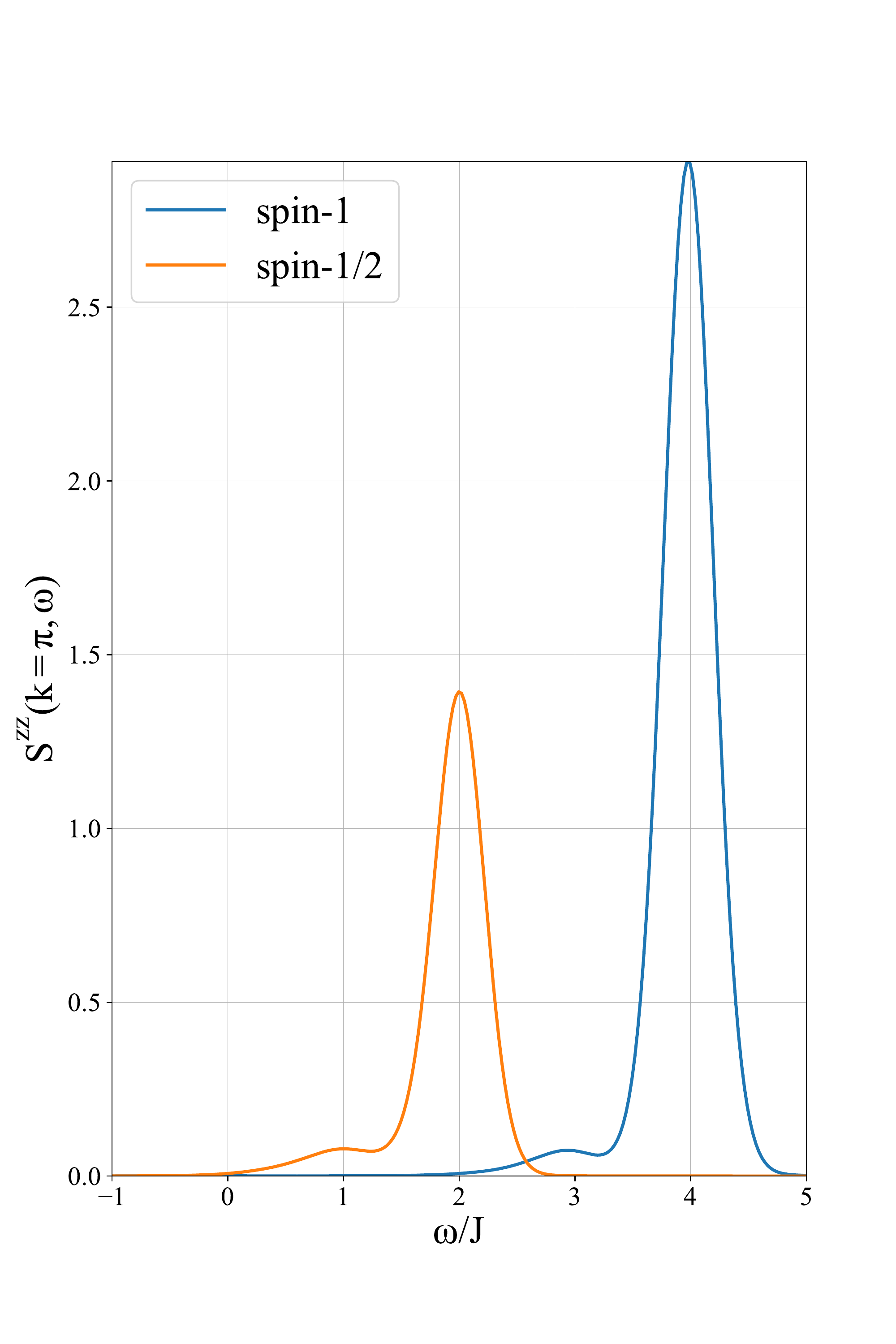}
\caption{Comparison between the thermal DSF section cut of the spin-1/2 and spin-1 FM Heisenberg chains at $k =\pi$ and $\beta J = 8$. In order to have an appropriate comparison we present here data obtained from keeping the same parameters for both chains, namely $L = 60$, $\chi = 400$ and $t_f = 14/J$. The height of the bound state peak is the same,  but the peak corresponding to the spin-wave excitation is twice as large for spin -1 than for spin-1/2 chain. }
\label{comparison_spin_half_spin1_chain}
\end{figure}

The first point of difference between spin-1/2 and spin-1 FM chains is that the main spin-wave excitation spectral peak is twice as large for the spin-1 chain as compared to the spin-1/2 chain (see Fig. \ref{comparison_spin_half_spin1_chain}). At the same time, the spectral peak corresponding to the bound state is of the same height for both cases. Therefore, the relative intensity of the bound state with respect to the main spin-wave is halved. Accordingly, detecting the bound state can be expected to be more challenging in spin-1 FM chains.\par 

\begin{figure}
\centering
\includegraphics[width =9.25cm, height=9.25cm , keepaspectratio]{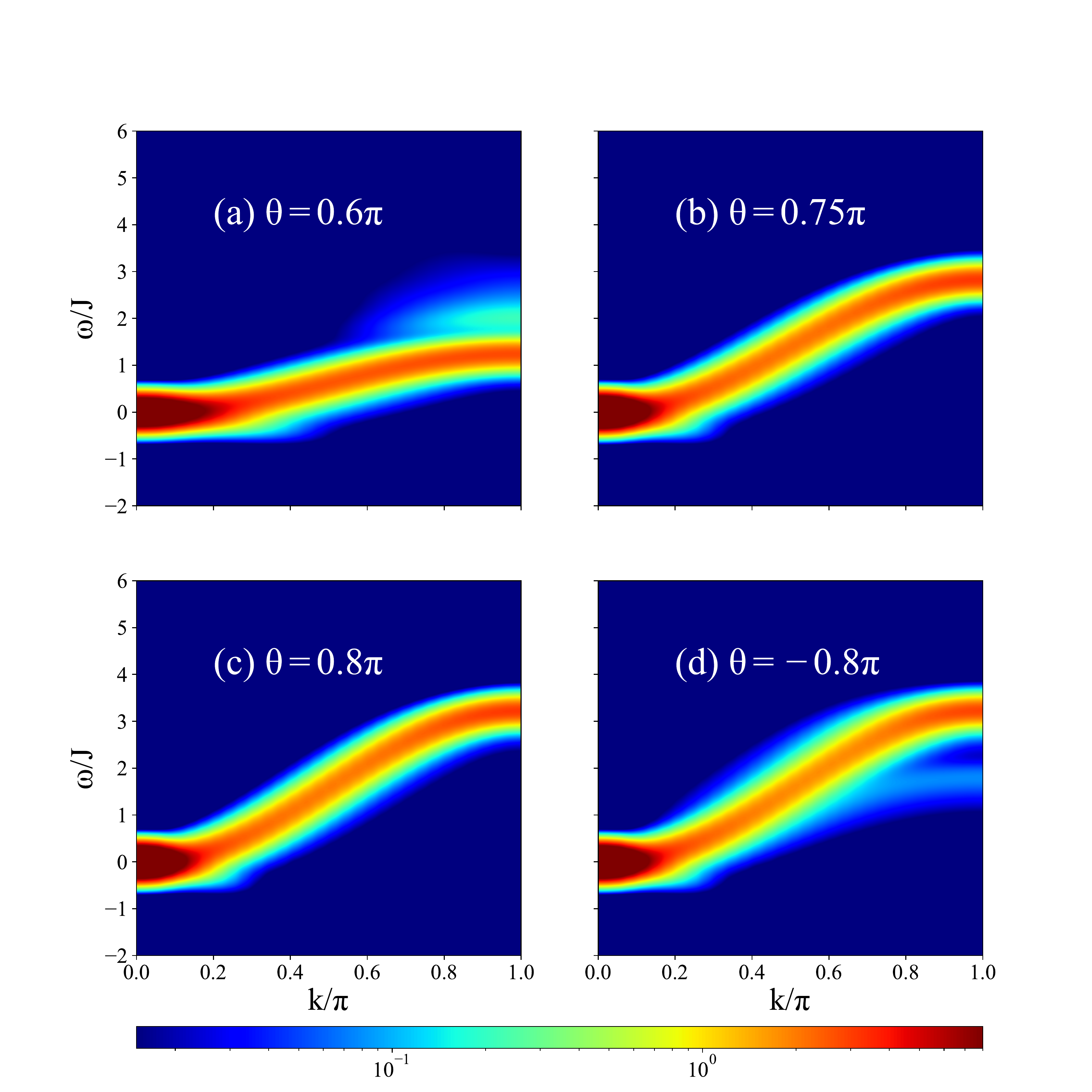}
\caption{Longitudinal thermal DSF ($S^{zz}(k,\omega)$) of the FM bilinear-biquadratic spin-1 chain ($L =60$, $T =J/10$, $t_f=16/J$). As the biquadratic coupling is decreased from (a) to (d), the clear signature of an anti-boundstate is replaced by that of a boundstate. Note that the anti-boundstate gathers very little spectral weight at finite temperature.}
\label{Thermal_DSF_BLBQ}
\end{figure}

For completeness, we also explored the thermal spectral signature of bound states and anti-bound states by thermal DSF simulation of the spin-1 BLBQ chain as shown in Fig. \ref{Thermal_DSF_BLBQ}. The anti-bound states lose most of the spectral weights as compared to the zero temperature simulation of the DQSF. The visible spectral weight lies close to the region $k\approx \pi$ for values of $\theta$ where the biquadratic coupling is large enough so that the anti-bound state is separated from the continuum by a gap (Fig. \ref{Thermal_DSF_BLBQ}a). In the zero temperature DQSF plots, as the biquadratic interaction is decreased in the range of $\theta\in\left\lbrack 3\pi/4,9\pi/10\right)$, there is either an anti-bound state or a bound state at a given $k$ vector. This points to a three-way competition between anti-bound state, bound state and two-magnon scattering state when we introduce temperature. The bound states are barely visible, even though they exist for $\theta \in \lbrack 3\pi/4, \pi)$ (see Fig. \ref{DQSF_BLBQ}), but upon decreasing the temperature to the appropriate level for $\theta$ greater than $\pi$, the bound state gathers enough weight to be distinguished from the main spin-wave excitation. Thus, a finite-temperature INS experiment can observe the bound state better in the presence of a significant negative biquadratic interaction. Although this is not the most generic situation in experimental realisations of spin-1 chains, negative biquadratic interactions can be generated in the presence of quasi-degenerate orbitals \cite{Mila_Zhang_2000}.\par

\subsection{Easy axis single-ion anisotropy} 
\subsubsection{Zero temperature dynamics}
\begin{figure}
\centering
\includegraphics[width = 9.25cm, height = 9.25cm ,keepaspectratio]{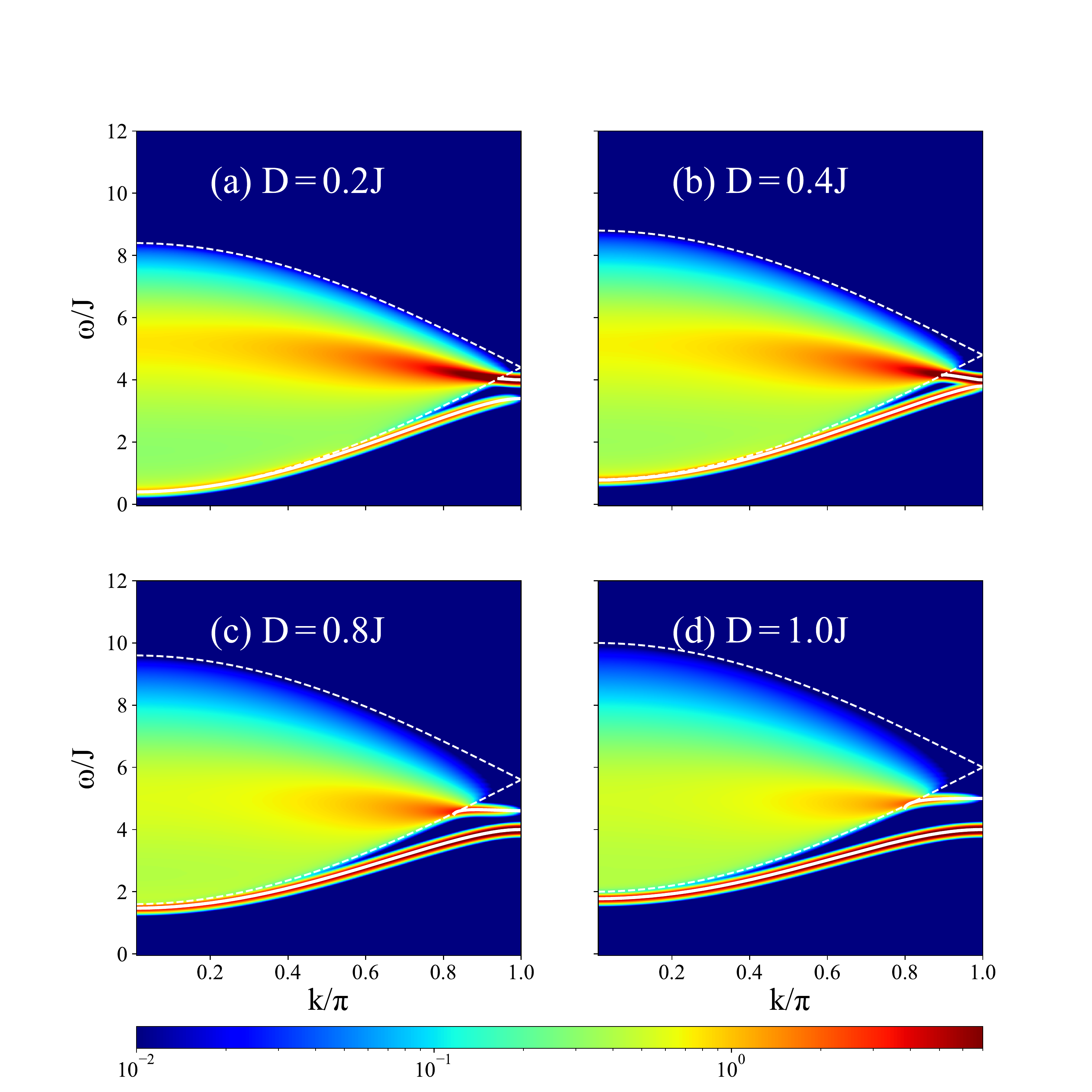}
\caption{Pairing component of the quadrupolar structure factor for spin-1 FM chain with different values of $D$ (L = 240, $t_f = 40/J$).  The bound state (denoted by solid lines) of two spin-waves already touches the continuum (denoted by dashed lines) at $k =\pi$ for $D=0.4J$. The resonance observed in the continuum is due to the single-ion bound state when $D<J/2$ and to the exchange bound state when $D>J/2$.}
\label{Fig_DQSF_values_D}  
\end{figure}

The spin-1 ferromagnetic chains are usually realised in Nickel based compounds and they exhibit an additional easy-axis single-ion anisotropy term (whose interaction strength is denoted by $D$) \cite{PChauhan_2020}. The Hamiltonian is given by : 
\begin{eqnarray}
H = -J\sum_{i}\mathbf{S}_{i}\cdot\mathbf{S}_{i+1}-D\sum_{i}{(S^z_i)}^2
\label{single_ion}
\end{eqnarray} 
The dispersion relation of the spin wave solution for the model is $2J(1-\cos k)+D$. In order to understand the modifications to the two spin-deviation excitation spectrum, we simulate the DQSF for this model. Because of the anisotropy, the three components are different from each other. For the longitudinal component ($\Delta S^z = 0$), the spectral weight is present only at $K=0$, $\omega = 0$ and in the transverse component ($\Delta S^z = 1$), only the single-spin-wave excitation has spectral weight. The pairing component ($\Delta S^z=2$) captures the spectral weights for (i) the two-spin-wave scattering states and (ii) the bound states.
\begin{enumerate}
\item The energies of the two-spin wave scattering states are shifted by $2D$ when compared to Heisenberg ferromagnet. 
\begin{eqnarray}
\omega_2(K,p) = 4J\left(1-\cos\frac{K}{2}\cos p\right)+2D,\nonumber
\end{eqnarray}
where the centre of mass momentum is denoted by $K$ and the relative momentum is denoted by $p$.
\item There are two bound states in this model corresponding to whether two-spin deviations are on the same sites or the adjacent sites. They are identified as single-ion spin bound state and exchange bound state respectively in the literature \cite{Silberglitt_Torrance_1970}. By following Wortis's method \cite{Wortis_1963,Silberglitt_Torrance_1970, Papanicolaou_1987} and setting up the transfer matrix equation, we arrive at two solutions for bound states. At $K =\pi$, the energy of single ion bound state is $4J$ and that of the exchange bound state is $3J+2D$. The difference in their energies varies as $J-2D$. There are two energy regimes  - for $D<J/2$, the single-ion spin bound state is at a higher energy and for $D>J/2$ the exchange bound state is at a higher energy. 
\end{enumerate}
The DQSF simulations in Fig. \ref{Fig_DQSF_values_D} show these excitations. It is interesting to note that the higher energy bound state enters the two-spin wave continuum and results in a resonance while the lower energy bound state exists as a separate mode completely (see Fig. \ref{Fig_DQSF_values_D}d)


\subsubsection{Finite temperature dynamics}
We explore the observability of bound states in the thermal DSF simulations. Due to anisotropy, the longitudinal and transverse component of DSF are different from each other.\par
\begin{figure}
\centering
\includegraphics[width= 9.25cm, height =9.25cm, keepaspectratio]{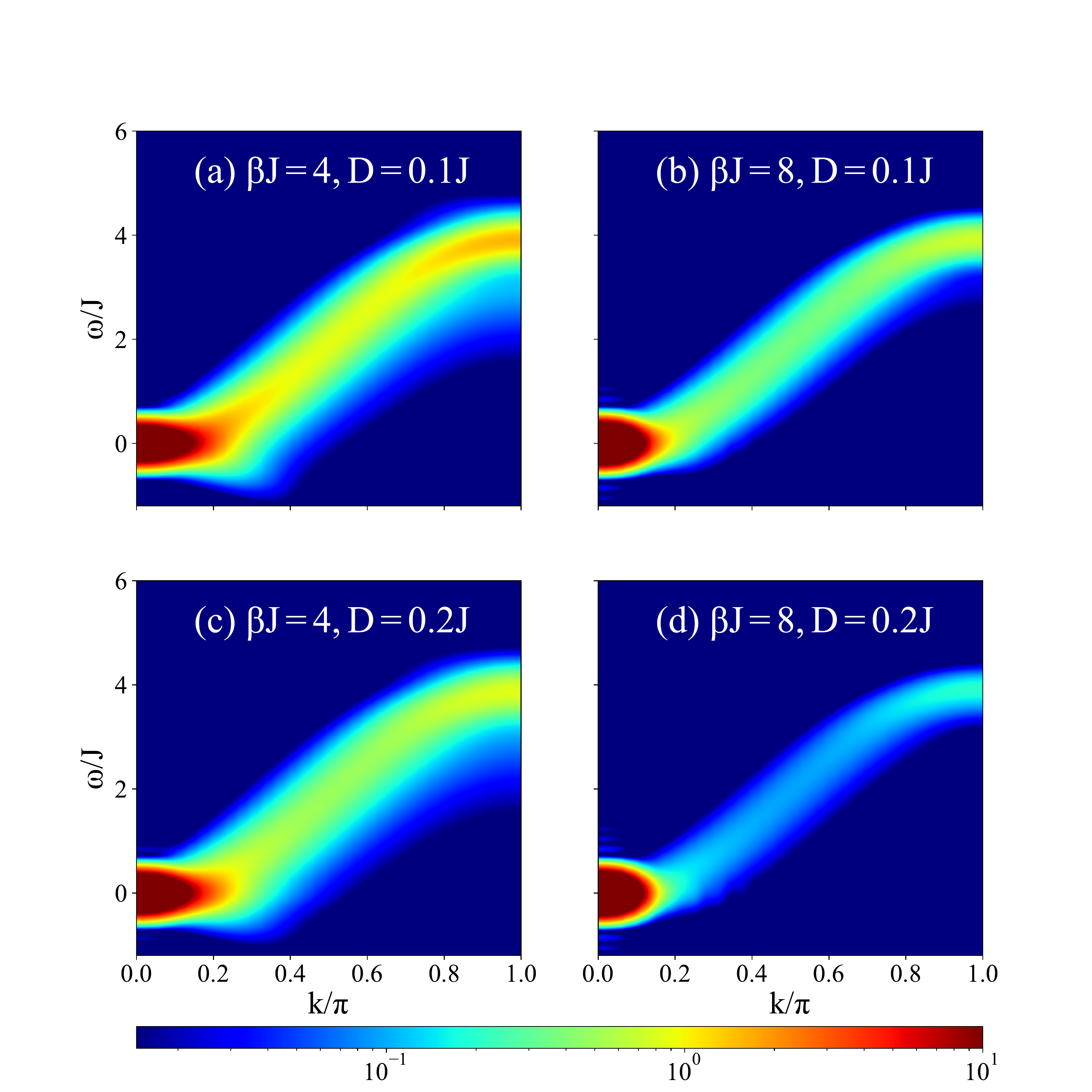}
\caption{Longitudinal component of the thermal DSF for the spin-1 FM chain with single-ion anisotropy (L =60, $t_f = 16/J$). Much of the spectral weight is concentrated at $k =0$ and $\omega = 0$. At higher temperatures and smaller magnitude of '$D$' the bound state gathers some spectral weight but it is indistinguishable from the thermally broadened spin-wave excitation.}
\label{Longitudinal_anisotropy_FM_spin1}
\end{figure}

(i) The longitudinal component shows most of the thermal spectral weights at $\omega=0$ and $k=0$ (Fig.\ref{Longitudinal_anisotropy_FM_spin1}).  At high temperatures, the spin-wave excitation gathers spectral weights and it remains unshifted by single-ion anisotropy $D$. For small anisotropy term, the exchange bound state gathers spectral weights as well, however due to thermal broadening it remains indistinguishable from the main spin-wave mode.\par
(ii) The transverse component of DSF has significant spectral weight along the dispersion relation of the spin-wave. Unlike the longitudinal component, the spin-wave excitation appears with shifted energies due to the single-ion anisotropy $D$. For high temperatures, there are spectral weights in the negative $\omega$-axis due to de-excitation processes from spin-wave state to fully-aligned state. For small $D$ values and at relatively higher temperatures, it captures spectral weights for the exchange bound state. The single-ion bound state decays via the continuum channel and does not gather significant spectral weights.

\begin{figure}
\centering
\includegraphics[width =9.25cm, height =9.25cm, keepaspectratio]{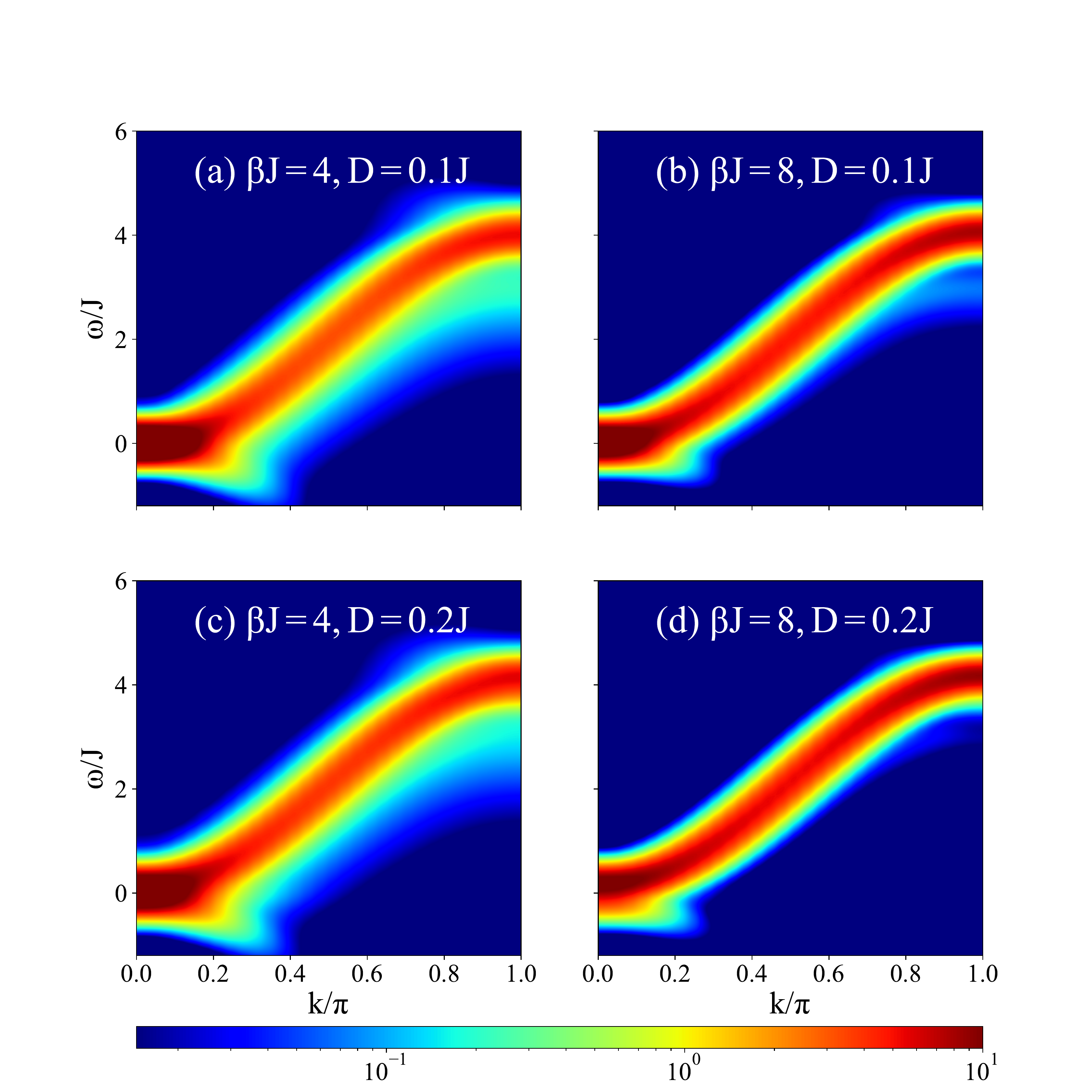}
\caption{Transverse component of the thermal DSF for the spin-1 FM chains with single-ion anisotropy (L = 60, $t_f = 16/J$). The spin-wave dispersion is shifted by $D$ as expected. The bound state is observable for smaller $D$ and higher temperature.}
\label{Transverse_anisotropy_FM_spin1}
\end{figure}

In zero temperature dynamics simulations, the gap between single-ion bound state and the exchange bound state reduces with increasing anisotropy.  Therefore, a more pronounced decay of the spectral weights for exchange bound states via the continuum channel is expected for larger values of the single-ion anisotropy ($D$).  So, the bound state can be expected to be seen in a ferromagnet with anisotropy $D<0.4J$. The spectral weights are exponentially suppressed as $e^{-\beta D}$.  As a result, the bound states are visible in the finite-temperature simulations at high temperatures (i.e. $\beta J=4$) and for small values of $D$ particularly within $0.2J$.  Due to thermal broadening of the bound state and the spin-wave, they are indistinguishable at higher temperatures. For larger anisotropy values i.e. $D>J/2$, the exchange bound state enters the continuum and therefore its spectral weight decays via the continuum channel. Although, in this regime, the single-ion bound state exists as a separate mode, its spectral weight is exponentially suppressed and it is not visible, even at higher temperatures. Our observations for this case are summarised in Fig.\ref{Transverse_anisotropy_FM_spin1}.


\subsection{Easy plane single-ion anisotropy} 

If an easy plane single-ion anisotropy is present instead, corresponding to $D<0$ in Eq. \ref{single_ion}, the ground state is no longer the fully aligned state but belongs to the sector $S^z_{tot}=0$, and the excitations are expected to have a very different nature. Since the present paper is devoted to the bound states of magnons, we do not discuss this case further.

\section{Conclusion}
The dynamics of multiple excitations in ferromagnets is very rich and it forms the bedrock to understand the antiferromagnetic case \cite{Bethe_1931,Toshiya_Inami1994, Elliott_1969,Headings_2010, Mourigal2013}. Upon diagonalising the two-spin deviation subspace of the 1D Heisenberg ferromagnet one finds the presence of a bound state. The thermal DMRG simulations reported here confirm that the bound state can be detected when the spin-wave states are thermally populated. 
From finite temperature numerics, we noted the long tail of the bound state spectral peak extending to very low energies and  found that the spectral weight of the bound state increases with temperature as $T^{\frac{3}{2}}$, a behaviour that is qualitatively captured by considering only contributions from prominent processes (i.e. excitations from fully aligned states and single spin deviation states).

INS experiments have already been conducted on ferromagnets, but at very low temperature (see e.g. Ref. \cite{Kopinga_1986}, where the exchange coupling ($J$) was 66 K while the measurements have been made at temperatures below 4K ($\sim J/16$). This temperature regime was not ideal to observe the bound state because its spectral weight is about one-hundredth of that of the spin-wave excitation in this temperature range. With the improvements in neutron scattering technology and performing a scan for $k=\pi$ (where the separation between the spin-wave and the bound state is maximum) at a higher temperature regime ($J/12$ to $J/3$), we have shown that it should be possible to detect the bound state in ferromagnetic chains, the most favourable case being spin-1/2. We also showed that including a magnetic field does not help since it tends to suppress the spectral weight corresponding to the bound state.

These arguments can in principle be extended to 2D and 3D Heisenberg ferromagnets where there are 2 bound states and 3 bound states at the edge of the Brillouin zones respectively, but differences are to be expected. In contrast to 1D, in 2D ferromagnets, there is one bound state which enters the continuum, while the other bound state exists for all k. In 3D, the three bound states only exist beyond certain threshold values of k close to the edge of the Brillouin zone. Thus, in higher dimensions, the bound states become more difficult to detect directly. Detecting the resonances where the bound state enters the continuum might be easier. Besides, the thermal broadening of the bound states and of the main spin-wave excitations might be additional bottlenecks in higher dimensions.\par

For higher values of the spin (like the spin-1 case studied in the present paper), the ratio of the difference between the bound state and the spin-wave energies and the main spin-wave energy at $k =\pi$ decreases. Thus, after thermal broadening, the bound state and the spin-wave peaks may not be easily distinguishable. Furthermore, the thermal DSF of spin-1 chains with easy-axis single ion anisotropy show that the bound state has less spectral weight and should be more difficult to detect than in the case with no anisotropy.
So the best candidate to observe the bound state is the spin-1/2 FM chain.

Finally, in real materials, finite temperature induces both acoustic and optical phonon modes which could obscure the bound state spectral weights in the magnon energy spectrum observed in INS experiments \cite{Jeske_2018, Tapan_Neutron_Scattering}. However, upon choosing a material with appropriate strength of magnetic exchange coupling, one should be able to separate the magnons from the phonons, especially at the edge of the Brillouin zone. Additionally, switching on the external magnetic field would decouple the spin degrees of freedom from the lattice vibrations and could lead to clear detection of the bound state.

In conclusion, we hope that the present paper will encourage specialists of inelastic neutron scattering to perform experiments at intermediate temperature to try and detect the bound state of the ferromagnetic chain, a ninety year old prediction still awaiting for a direct confirmation. More generally, beyond the case of the bound state of ferromagnets, the present results suggest that performing INS at intermediate temperature might help revealing excitations not visible at zero temperature.

\section*{Acknowledgements}
We are very grateful to Noam Kestin for insightful discussions on the Thermal DMRG code, and to Henrik R\o{}nnow for discussions regarding neutron scattering experiments. MN thanks Aubry Jaquier for helping in setting up the ALPS package. MN also thanks Olivier Gauth\'e and Jeanne Colbois for helpful discussions. The numerical simulations were performed on the SCITAS clusters at EPFL. We acknowledge the funding from the Swiss National Science Foundation. 
\appendix
\section{Two-spin deviation spectrum of the spin-1/2 FM chain}
\label{Appendix_spin_half_FM_chain}

\begin{figure}
\includegraphics[width =12cm ,height =12cm, keepaspectratio]{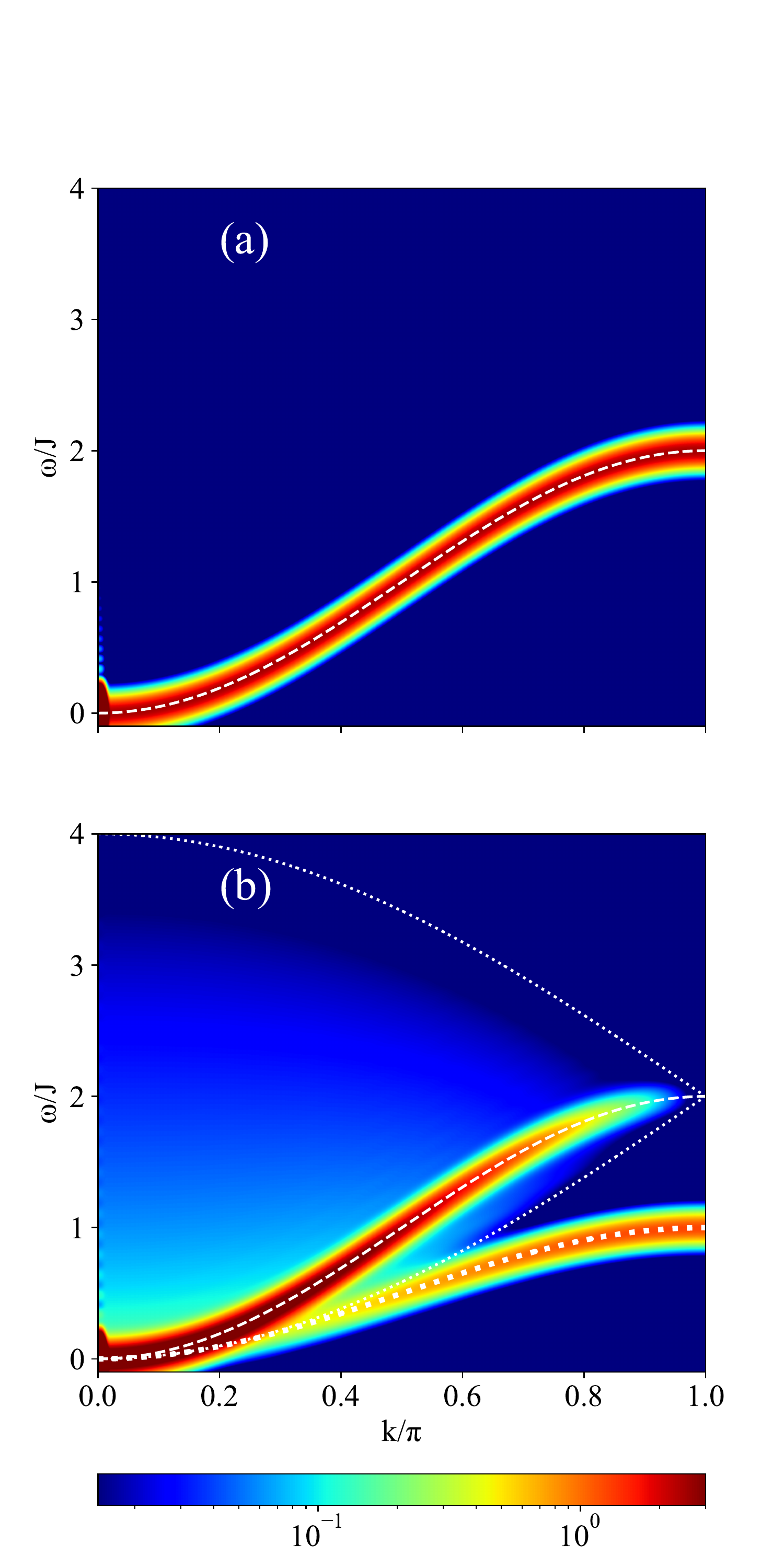}
\caption{Zero temperature structure factors of the one and two-spin deviation spectrum of the FM spin-$1/2$ chain with $J=1$, $L =240$ and $t_f=40/J$. (a) $S^{-,+}(k,\omega)$, the transverse component of DSF measured in INS experiments. It only shows the spin-wave excitation (white dashed lines); (b) $S_2(k,\omega)$, the two-spin deviation spectrum. White dashed lines: single spin wave; white dotted lines: two spin-wave continuum. The bound state of spin-waves is denoted by a thick white dotted line. Note that this structure factor is not measurable in an INS experiment.}
\label{zero_temp_spin_half}
\end{figure}

As discussed in the main text, the two-spin-deviation subspace of the FM Heisenberg model has two types of solutions: (i) two-spin-wave scattering states and (ii) a bound state of spin-waves. In this section, we discuss the form of the bound state.  One can consider a general state in the two-spin deviation subspace \cite{Fukuda_Wortis_1963, DCMattis, Keselman_2020} as:
\begin{eqnarray}
{|2\rangle}_{k_1,k_2} = \sum_{x_1< x_2} a_{k_1,k_2}(x_1,x_2) S^{+}_{x_1}S^{+}_{x_2}|\mathrm{GS}\rangle 
\label{two_spin_wave_state}
\end{eqnarray}
where $a_{k_1,k_2}(x_1,x_2)$ is the coefficient dependent on the position of the spin-deviations in the chain. One can assume the Bethe-Ansatz form of the solution, i.e. superposition of an incident wave and a scattered wave. One can work in the centre of mass coordinates defined by
\begin{eqnarray}
&K = k_1 + k_2,\hspace{1cm} &R_{12} = \frac{r_{x_1}+r_{x_2}}{2}\nonumber\\
&p = \frac{k_1 - k_2}{2},\hspace{1cm} &r_{12} = r_{x_1}-r_{x_2}\nonumber
\end{eqnarray}
to get the following results:
\begin{enumerate}
\item For the two-spin-wave scattering states, one can proceed to develop the eigenvalue equation and to find the travelling solution
$$a_{K,p} \approx  \frac{\sqrt{2}}{L}\cos\lbrack pr_{12}+\theta \rbrack e^{iKR_{12}}$$
where the definition of the phase factor $\theta$ is obtained from the boundary condition of the transfer matrix equation as follows:
\begin{equation}
\cot \theta = \frac{\sin p}{\cos\frac{K}{2} - \cos p} \nonumber
\end{equation}
\item For the bound state, one gets
\begin{equation}
a_{K,\tilde{p}}(x_1,x_2)= \frac{1}{2\sqrt{L}}\frac{|\sin\frac{K}{2}|}{\sqrt{1-\cos^{L}\frac{K}{2}}}e^{iKR_{12}}e^{-\tilde{p}(|r_{12}|-1)}
\label{coeff_bound_state}
\end{equation}
where, $\tilde{p}$ is defined as 
\begin{equation}
\tilde{p} = -\ln \cos \frac{K}{2}\nonumber
\label{imp_reln_bc}
\end{equation}
\end{enumerate}
Numerically, the transverse dynamical structure factor of a spin-1/2 FM chain at zero temperature (i.e. $\beta\to\infty$) is determined by evaluating Eq. \ref{Spin_DSF_def} with $\alpha=-$, $\tilde{\alpha}=+$. This leads to:
\begin{eqnarray}
S^{-,+}(k,\omega) &=& \frac{1}{L}\sum_{i,j} e^{-ik (r_j-r_i)}e^{i\omega t}\langle \mathrm{GS}| S^{-}(j,t)S^{+}(i,0)|\mathrm{GS}\rangle\nonumber\\
&=& \frac{2\pi}{L}\delta(\omega-\omega_1(k))
\end{eqnarray}
So, in an INS experiment, one just observes the spin-wave excitation (shown in Fig. \ref{zero_temp_spin_half}a) whose dispersion relation $\omega_1(k)$ is given in Eq.\ref{spin12_spinwave} of the main text. 

To probe the two-spin deviation sector, it is convenient to study operators which lead to two-spin deviations in the FM ground state, and to use the time-dependent DMRG algorithm to compute the time-dependent correlation functions based on these operators. 

One such operator and its conjugate are defined by: 
\begin{equation}
\hat{O}^{\pm}_{2}(i) =  S^{\pm}(i)S^{\pm}(i+1)\nonumber
\end{equation}
and the two-spin-deviation correlation function reads
\begin{equation}
C_{2}(i,j;t) = \frac{1}{2}\left\lbrack\bra{\mathrm{GS}}\hat{O}^{-}_{2}(j,t)\hat{O}^{+}_{2}(i,0)\ket{\mathrm{GS}}+\mathrm{h.c.}\right\rbrack\nonumber
\end{equation}
The two-spin wave dynamical structure factor is given by the space and time Fourier transform of the correlation function: 
\begin{equation}
S_2(k,\omega) = \frac{1}{L^2}\int_{-\infty}^{\infty}dt\sum_{i,j}e^{-ik (r_j-r_i)}e^{i\omega t}C_2(i,j;t)\nonumber
\end{equation}
The numerical evaluation of $S_2(k,\omega)$ clearly shows a two-spin-wave continuum and a bound state of spin-waves, as shown in Fig. \ref{zero_temp_spin_half}b. This confirms the structure of the two spin deviation subspace. However, we are not aware of any experimental technique in which a 2-spin-wave spectrum can be measured at very low temperatures for the spin chain compounds.\par

\section{Numerical details}
\subsection{Determination of the temperature range}
\label{temp_range}
In this section, we explain the criteria we used to choose the temperature range in which the bound state can be observed in the thermal DSF. For determining the lower limit on the temperature, we compared the area under the bound state spectral peak at $k=\pi$ for various temperatures. As the temperature lowers, there is less spectral weight in the bound state ($I_{\mathrm{BS}}$) as compared to the dominant spin-wave feature ($I_{\mathrm{SW}}$). In order to limit the finite size effects, the area under the spectral peak for various sizes of the chain were extrapolated in the thermodynamic limit. In Table \ref{Intensity_table}, we show the data obtained from the formula $I_{\mathrm{BS}}/(I_{\mathrm{BS}}+I_{\mathrm{SW}})$. In terms of percentage of total spectral weights, we consider the cut off temperature at which the bound state is detectable to be 5 \%. This leads to the lower limit  $k_BT  > J/12$.\par
\begin{figure}
\centering
\includegraphics[width=9cm, height =9cm, keepaspectratio]{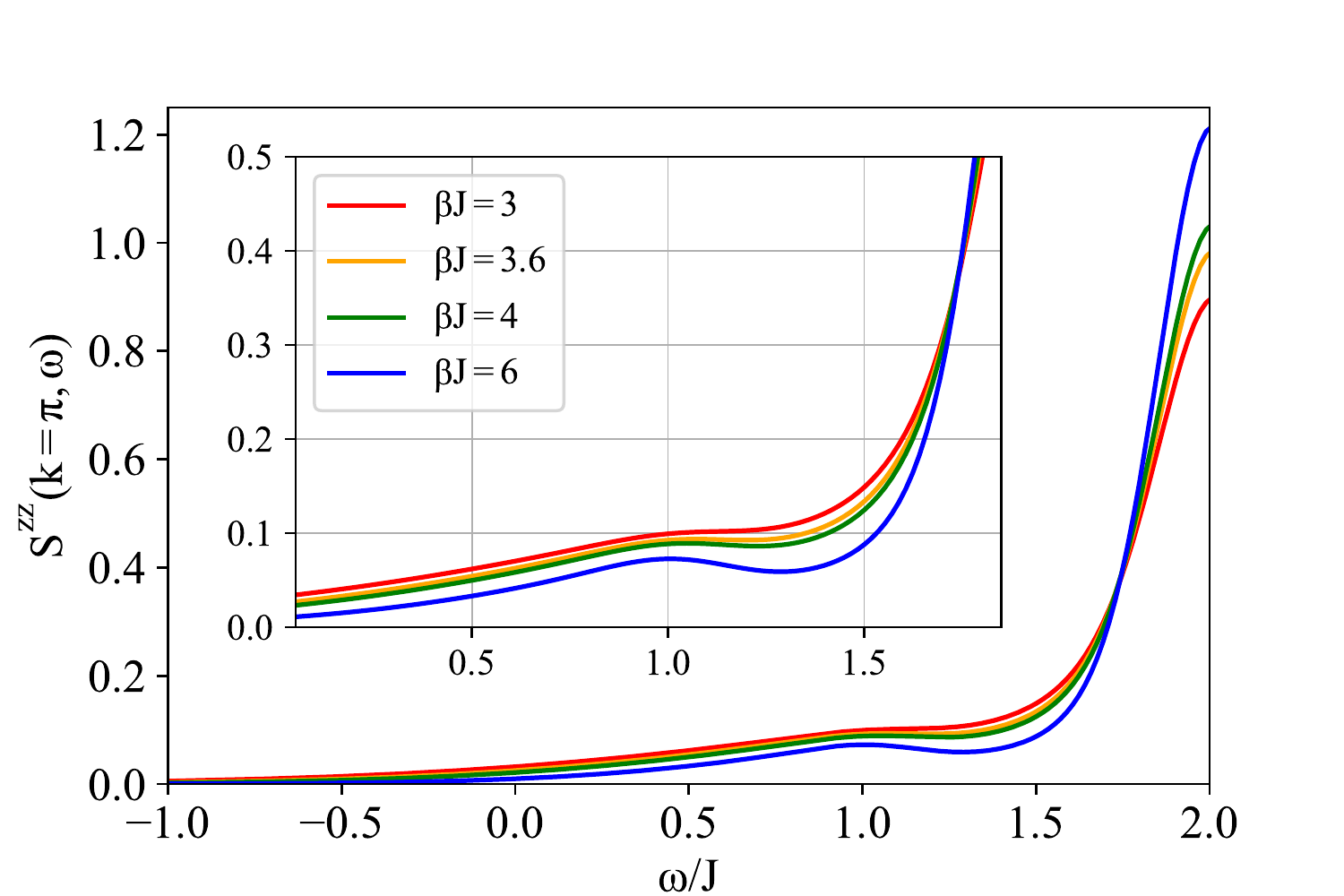}
\caption{Section cut of longitudinal thermal DSF at $k=\pi$ of the spin-1/2 FM chain. The spectral peak of the bound state is no longer distinguishable from the spectral peak of the main spin-wave mode due to thermal broadening. Inset: zoomed-in picture used to determine the temperature $T = J/3$ below which we still get two separate peaks.}
\label{longitudinal_DSF_section_cut}
\end{figure}

\begin{table}
\begin{tabular}{|c|c|c|c|}
\hline
 $\beta J$ & $I_{\mathrm{BS}}(k=\pi,\omega)$ & $I_{\mathrm{SW}}(k=\pi,\omega)$ & $\frac{I_{\mathrm{BS}}}{I_{\mathrm{BS}}+I_{\mathrm{SW}}}\times100$\\
 \hline
$4$ &0.0801 & 0.5081 &13.620\\
$6$ &0.05846 & 0.53191 &9.902\\
$8$ &0.04391 & 0.54691&7.432\\
$10$ &0.0348 & 0.5561 &5.890\\
$12$ &0.0283 & 0.5626 &4.790\\
\hline
\end{tabular}
\caption{Comparison of the areas under the curve associated with the bound state and with the main spin-wave excitation at various temperatures for the spin-1/2 FM chain. The percentage of spectral weight corresponding to the bound state with respect to the total spectral weight at $k=\pi$ is used as a criterion for the bound state to be observable.}
\label{Intensity_table}
\end{table}
The upper limit on the temperature is set by the thermal broadening of the modes, as illustrated in Fig.\ref{longitudinal_DSF_section_cut}. Above a certain temperature, the minimum between the peaks of the bound state and of the magnon disappears, and the peak of the bound state becomes a shoulder of the magnon peak. This temperature corresponds to the upper limit quoted in the main text, $k_BT<J/3$. 

\subsection{Finite size analysis of thermodynamic quantities}
\label{finite_size_scaling_therm_quant}
\begin{figure}
\centering
\includegraphics[width=9cm,height=9cm,keepaspectratio]{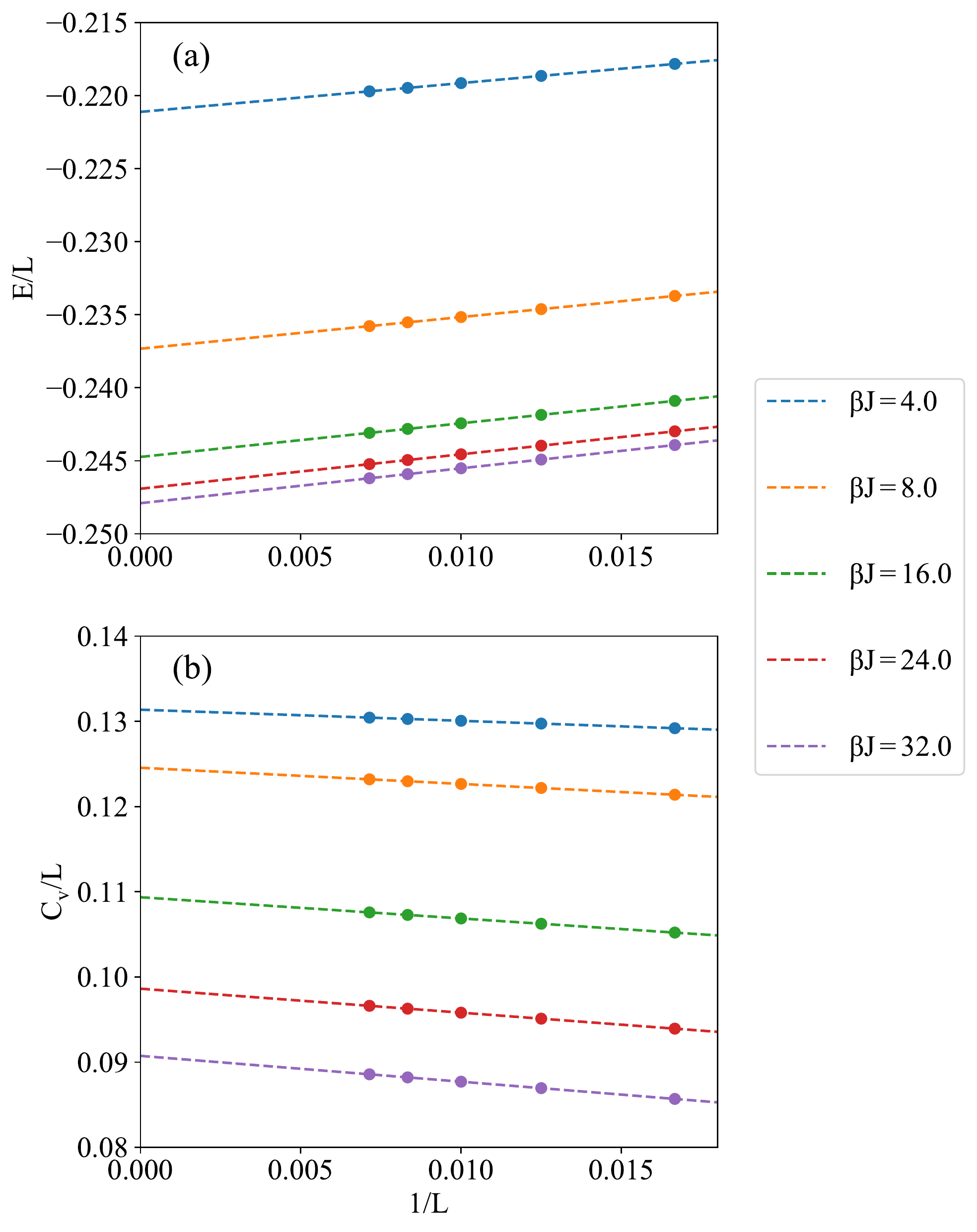}
\caption{Ensemble energy per unit length (a) and specific heat per unit length (b) for the spin-1/2 Heisenberg FM chain extrapolated linearly with respect to $1/L$ for a few temperatures.  
}
\label{extrapolation_thermal_quantities}
\end{figure}

To perform a finite-size scaling analysis of the  average energy($\langle E\rangle/L$) and of the specific heat ($C_v/L$), we have considered chains of length L=60, 80, 100, 120 and 140. These data can be extrapolated linearly with respect to $1/L$ with very good accuracy at every temperature, as shown in Fig. \ref{extrapolation_thermal_quantities}. The entropy was determined by integrating the already extrapolated specific heat and similarly the free energy was obtained from the extrapolated  energy and entropy. We used these data for the plots in Fig. \ref{Thermal_statics_FM} of the main text.
\subsection{Finite size analysis of spectral weights}
\label{finite_size_scaling_bound_state}
\begin{figure}
\centering
\includegraphics[width=9cm, height=9cm, keepaspectratio]{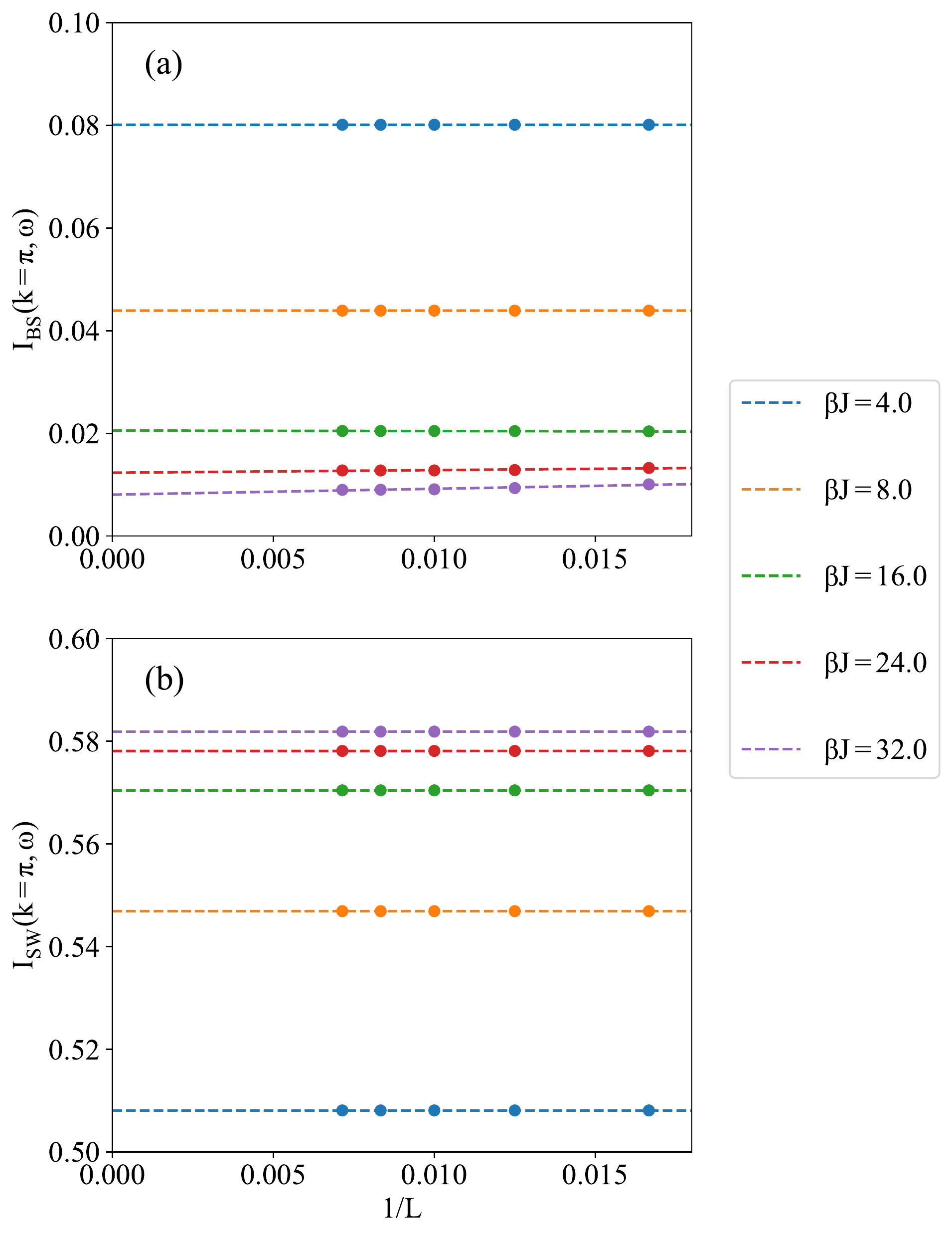}
\caption{Area under the curve corresponding to the bound state (a) and to the main spin-wave mode (b) extrapolated linearly with respect to $1/L$ for a few temperatures. The finite-size effects are very small.}
\label{extrapolation_bound_state_spin_wave_area}
\end{figure}

In order to plot Fig. \ref{section_cut_area_comparison}b in the main text, we determined the area corresponding to the bound state ($I_{\mathrm{BS}}$) and the main spin-wave mode ($I_{\mathrm{SW}}$) of the section cut at $k=\pi$ for various sizes, and we performed a linear extrapolation in $1/L$ to the thermodynamic limit. We also plot the extrapolation data for main spin-wave mode in Fig. \ref{extrapolation_bound_state_spin_wave_area} for completeness. In all cases, the finite-size effects are very small. 

\bibliography{references.bib}

\begin{thebibliography}{45}%
\makeatletter
\providecommand \@ifxundefined [1]{%
 \@ifx{#1\undefined}
}%
\providecommand \@ifnum [1]{%
 \ifnum #1\expandafter \@firstoftwo
 \else \expandafter \@secondoftwo
 \fi
}%
\providecommand \@ifx [1]{%
 \ifx #1\expandafter \@firstoftwo
 \else \expandafter \@secondoftwo
 \fi
}%
\providecommand \natexlab [1]{#1}%
\providecommand \enquote  [1]{``#1''}%
\providecommand \bibnamefont  [1]{#1}%
\providecommand \bibfnamefont [1]{#1}%
\providecommand \citenamefont [1]{#1}%
\providecommand \href@noop [0]{\@secondoftwo}%
\providecommand \href [0]{\begingroup \@sanitize@url \@href}%
\providecommand \@href[1]{\@@startlink{#1}\@@href}%
\providecommand \@@href[1]{\endgroup#1\@@endlink}%
\providecommand \@sanitize@url [0]{\catcode `\\12\catcode `\$12\catcode
  `\&12\catcode `\#12\catcode `\^12\catcode `\_12\catcode `\%12\relax}%
\providecommand \@@startlink[1]{}%
\providecommand \@@endlink[0]{}%
\providecommand \url  [0]{\begingroup\@sanitize@url \@url }%
\providecommand \@url [1]{\endgroup\@href {#1}{\urlprefix }}%
\providecommand \urlprefix  [0]{URL }%
\providecommand \Eprint [0]{\href }%
\providecommand \doibase [0]{http://dx.doi.org/}%
\providecommand \selectlanguage [0]{\@gobble}%
\providecommand \bibinfo  [0]{\@secondoftwo}%
\providecommand \bibfield  [0]{\@secondoftwo}%
\providecommand \translation [1]{[#1]}%
\providecommand \BibitemOpen [0]{}%
\providecommand \bibitemStop [0]{}%
\providecommand \bibitemNoStop [0]{.\EOS\space}%
\providecommand \EOS [0]{\spacefactor3000\relax}%
\providecommand \BibitemShut  [1]{\csname bibitem#1\endcsname}%
\let\auto@bib@innerbib\@empty
\bibitem [{\citenamefont {Bethe}(1931)}]{Bethe_1931}%
  \BibitemOpen
  \bibfield  {author} {\bibinfo {author} {\bibfnamefont {H.}~\bibnamefont
  {Bethe}},\ }\href@noop {} {\bibfield  {journal} {\bibinfo  {journal} {Journal
  Zeitschrift für Physik}\ }\textbf {\bibinfo {volume} {71}},\ \bibinfo
  {pages} {205} (\bibinfo {year} {1931})}\BibitemShut {NoStop}%
\bibitem [{\citenamefont {Bloch}(1930)}]{Bloch}%
  \BibitemOpen
  \bibfield  {author} {\bibinfo {author} {\bibfnamefont {F.}~\bibnamefont
  {Bloch}},\ }\href {https://doi.org/10.1007/BF01339661} {\bibfield  {journal}
  {\bibinfo  {journal} {Zeitschrift für Physik}\ }\textbf {\bibinfo {volume}
  {61}},\ \bibinfo {pages} {206} (\bibinfo {year} {1930})}\BibitemShut
  {NoStop}%
\bibitem [{\citenamefont {Dyson}(1956)}]{Dyson_1956}%
  \BibitemOpen
  \bibfield  {author} {\bibinfo {author} {\bibfnamefont {F.~J.}\ \bibnamefont
  {Dyson}},\ }\href {\doibase 10.1103/PhysRev.102.1217} {\bibfield  {journal}
  {\bibinfo  {journal} {Phys. Rev.}\ }\textbf {\bibinfo {volume} {102}},\
  \bibinfo {pages} {1217} (\bibinfo {year} {1956})}\BibitemShut {NoStop}%
\bibitem [{\citenamefont {Wortis}(1963)}]{Wortis_1963}%
  \BibitemOpen
  \bibfield  {author} {\bibinfo {author} {\bibfnamefont {M.}~\bibnamefont
  {Wortis}},\ }\href {\doibase 10.1103/PhysRev.132.85} {\bibfield  {journal}
  {\bibinfo  {journal} {Phys. Rev.}\ }\textbf {\bibinfo {volume} {132}},\
  \bibinfo {pages} {85} (\bibinfo {year} {1963})}\BibitemShut {NoStop}%
\bibitem [{\citenamefont {Mattis}(1981)}]{DCMattis}%
  \BibitemOpen
  \bibfield  {author} {\bibinfo {author} {\bibfnamefont {D.~C.}\ \bibnamefont
  {Mattis}},\ }\href@noop {} {\emph {\bibinfo {title} {The Theory of Magnetism
  I}}}\ (\bibinfo  {publisher} {Springer},\ \bibinfo {year} {1981})\BibitemShut
  {NoStop}%
\bibitem [{\citenamefont {Haldane}(1982{\natexlab{a}})}]{Haldane_1982a}%
  \BibitemOpen
  \bibfield  {author} {\bibinfo {author} {\bibfnamefont {F.~D.~M.}\
  \bibnamefont {Haldane}},\ }\href {\doibase 10.1088/0022-3719/15/36/008}
  {\bibfield  {journal} {\bibinfo  {journal} {Journal of Physics C: Solid State
  Physics}\ }\textbf {\bibinfo {volume} {15}},\ \bibinfo {pages} {L1309}
  (\bibinfo {year} {1982}{\natexlab{a}})}\BibitemShut {NoStop}%
\bibitem [{\citenamefont {Haldane}(1982{\natexlab{b}})}]{Haldane_1982b}%
  \BibitemOpen
  \bibfield  {author} {\bibinfo {author} {\bibfnamefont {F.~D.~M.}\
  \bibnamefont {Haldane}},\ }\href {\doibase 10.1088/0022-3719/15/24/006}
  {\bibfield  {journal} {\bibinfo  {journal} {Journal of Physics C: Solid State
  Physics}\ }\textbf {\bibinfo {volume} {15}},\ \bibinfo {pages} {L831}
  (\bibinfo {year} {1982}{\natexlab{b}})}\BibitemShut {NoStop}%
\bibitem [{\citenamefont {Mikeska}(1977)}]{Mikeska_1977}%
  \BibitemOpen
  \bibfield  {author} {\bibinfo {author} {\bibfnamefont {H.~J.}\ \bibnamefont
  {Mikeska}},\ }\href {\doibase 10.1088/0022-3719/11/1/007} {\bibfield
  {journal} {\bibinfo  {journal} {Journal of Physics C: Solid State Physics}\
  }\textbf {\bibinfo {volume} {11}},\ \bibinfo {pages} {L29} (\bibinfo {year}
  {1977})}\BibitemShut {NoStop}%
\bibitem [{\citenamefont {Bohn}\ \emph {et~al.}(1980)\citenamefont {Bohn},
  \citenamefont {Zinn}, \citenamefont {Dorner},\ and\ \citenamefont
  {Kollmar}}]{Bohn_1980}%
  \BibitemOpen
  \bibfield  {author} {\bibinfo {author} {\bibfnamefont {H.~G.}\ \bibnamefont
  {Bohn}}, \bibinfo {author} {\bibfnamefont {W.}~\bibnamefont {Zinn}}, \bibinfo
  {author} {\bibfnamefont {B.}~\bibnamefont {Dorner}}, \ and\ \bibinfo {author}
  {\bibfnamefont {A.}~\bibnamefont {Kollmar}},\ }\href {\doibase
  10.1103/PhysRevB.22.5447} {\bibfield  {journal} {\bibinfo  {journal} {Phys.
  Rev. B}\ }\textbf {\bibinfo {volume} {22}},\ \bibinfo {pages} {5447}
  (\bibinfo {year} {1980})}\BibitemShut {NoStop}%
\bibitem [{\citenamefont {Kopinga}\ \emph {et~al.}(1986)\citenamefont
  {Kopinga}, \citenamefont {de~Jonge}, \citenamefont {Steiner}, \citenamefont
  {de~Vries},\ and\ \citenamefont {Frikkee}}]{Kopinga_1986}%
  \BibitemOpen
  \bibfield  {author} {\bibinfo {author} {\bibfnamefont {K.}~\bibnamefont
  {Kopinga}}, \bibinfo {author} {\bibfnamefont {W.~J.~M.}\ \bibnamefont
  {de~Jonge}}, \bibinfo {author} {\bibfnamefont {M.}~\bibnamefont {Steiner}},
  \bibinfo {author} {\bibfnamefont {G.~C.}\ \bibnamefont {de~Vries}}, \ and\
  \bibinfo {author} {\bibfnamefont {E.}~\bibnamefont {Frikkee}},\ }\href
  {\doibase 10.1103/PhysRevB.34.4826} {\bibfield  {journal} {\bibinfo
  {journal} {Phys. Rev. B}\ }\textbf {\bibinfo {volume} {34}},\ \bibinfo
  {pages} {4826} (\bibinfo {year} {1986})}\BibitemShut {NoStop}%
\bibitem [{\citenamefont {Chauhan}\ \emph {et~al.}(2020)\citenamefont
  {Chauhan}, \citenamefont {Mahmood}, \citenamefont {Changlani}, \citenamefont
  {Koohpayeh},\ and\ \citenamefont {Armitage}}]{PChauhan_2020}%
  \BibitemOpen
  \bibfield  {author} {\bibinfo {author} {\bibfnamefont {P.}~\bibnamefont
  {Chauhan}}, \bibinfo {author} {\bibfnamefont {F.}~\bibnamefont {Mahmood}},
  \bibinfo {author} {\bibfnamefont {H.~J.}\ \bibnamefont {Changlani}}, \bibinfo
  {author} {\bibfnamefont {S.~M.}\ \bibnamefont {Koohpayeh}}, \ and\ \bibinfo
  {author} {\bibfnamefont {N.~P.}\ \bibnamefont {Armitage}},\ }\href {\doibase
  10.1103/PhysRevLett.124.037203} {\bibfield  {journal} {\bibinfo  {journal}
  {Phys. Rev. Lett.}\ }\textbf {\bibinfo {volume} {124}},\ \bibinfo {pages}
  {037203} (\bibinfo {year} {2020})}\BibitemShut {NoStop}%
\bibitem [{\citenamefont {Chatterjee}(2006)}]{Tapan_Neutron_Scattering}%
  \BibitemOpen
  \bibfield  {author} {\bibinfo {author} {\bibfnamefont {T.}~\bibnamefont
  {Chatterjee}},\ }\href@noop {} {\emph {\bibinfo {title} {Neutron Scattering
  from Magnetic Materials}}}\ (\bibinfo  {publisher} {Elsevier},\ \bibinfo
  {year} {2006})\BibitemShut {NoStop}%
\bibitem [{\citenamefont {{De Vries}}\ \emph {et~al.}(1989)\citenamefont {{De
  Vries}}, \citenamefont {Frikkee}, \citenamefont {Kakurai}, \citenamefont
  {Steiner}, \citenamefont {Dorner}, \citenamefont {Kopinga},\ and\
  \citenamefont {{De Jonge}}}]{DeVries_1989}%
  \BibitemOpen
  \bibfield  {author} {\bibinfo {author} {\bibfnamefont {G.}~\bibnamefont {{De
  Vries}}}, \bibinfo {author} {\bibfnamefont {E.}~\bibnamefont {Frikkee}},
  \bibinfo {author} {\bibfnamefont {K.}~\bibnamefont {Kakurai}}, \bibinfo
  {author} {\bibfnamefont {M.}~\bibnamefont {Steiner}}, \bibinfo {author}
  {\bibfnamefont {B.}~\bibnamefont {Dorner}}, \bibinfo {author} {\bibfnamefont
  {K.}~\bibnamefont {Kopinga}}, \ and\ \bibinfo {author} {\bibfnamefont
  {W.}~\bibnamefont {{De Jonge}}},\ }\href {\doibase
  https://doi.org/10.1016/0921-4526(89)90649-2} {\bibfield  {journal} {\bibinfo
   {journal} {Physica B: Condensed Matter}\ }\textbf {\bibinfo {volume}
  {156-157}},\ \bibinfo {pages} {266} (\bibinfo {year} {1989})}\BibitemShut
  {NoStop}%
\bibitem [{\citenamefont {Silberglitt}\ and\ \citenamefont
  {Harris}(1967)}]{Silberglit_Harris_1967}%
  \BibitemOpen
  \bibfield  {author} {\bibinfo {author} {\bibfnamefont {R.}~\bibnamefont
  {Silberglitt}}\ and\ \bibinfo {author} {\bibfnamefont {A.~B.}\ \bibnamefont
  {Harris}},\ }\href {\doibase 10.1103/PhysRevLett.19.30} {\bibfield  {journal}
  {\bibinfo  {journal} {Phys. Rev. Lett.}\ }\textbf {\bibinfo {volume} {19}},\
  \bibinfo {pages} {30} (\bibinfo {year} {1967})}\BibitemShut {NoStop}%
\bibitem [{\citenamefont {Silberglitt}\ and\ \citenamefont
  {Harris}(1968)}]{Silberglitt_Harris_1968}%
  \BibitemOpen
  \bibfield  {author} {\bibinfo {author} {\bibfnamefont {R.}~\bibnamefont
  {Silberglitt}}\ and\ \bibinfo {author} {\bibfnamefont {A.~B.}\ \bibnamefont
  {Harris}},\ }\href {\doibase 10.1103/PhysRev.174.640} {\bibfield  {journal}
  {\bibinfo  {journal} {Phys. Rev.}\ }\textbf {\bibinfo {volume} {174}},\
  \bibinfo {pages} {640} (\bibinfo {year} {1968})}\BibitemShut {NoStop}%
\bibitem [{\citenamefont {Date}\ and\ \citenamefont
  {Motokawa}(1966)}]{Date_Motokawa_1966}%
  \BibitemOpen
  \bibfield  {author} {\bibinfo {author} {\bibfnamefont {M.}~\bibnamefont
  {Date}}\ and\ \bibinfo {author} {\bibfnamefont {M.}~\bibnamefont
  {Motokawa}},\ }\href {\doibase 10.1103/PhysRevLett.16.1111} {\bibfield
  {journal} {\bibinfo  {journal} {Phys. Rev. Lett.}\ }\textbf {\bibinfo
  {volume} {16}},\ \bibinfo {pages} {1111} (\bibinfo {year}
  {1966})}\BibitemShut {NoStop}%
\bibitem [{\citenamefont {Torrance}\ and\ \citenamefont
  {Tinkham}(1969)}]{Torrance_Tinkham_1969}%
  \BibitemOpen
  \bibfield  {author} {\bibinfo {author} {\bibfnamefont {J.~B.}\ \bibnamefont
  {Torrance}}\ and\ \bibinfo {author} {\bibfnamefont {M.}~\bibnamefont
  {Tinkham}},\ }\href {\doibase 10.1103/PhysRev.187.595} {\bibfield  {journal}
  {\bibinfo  {journal} {Phys. Rev.}\ }\textbf {\bibinfo {volume} {187}},\
  \bibinfo {pages} {595} (\bibinfo {year} {1969})}\BibitemShut {NoStop}%
\bibitem [{\citenamefont {Barthel}\ \emph {et~al.}(2009)\citenamefont
  {Barthel}, \citenamefont {Schollw\"ock},\ and\ \citenamefont
  {White}}]{Barthel_2009}%
  \BibitemOpen
  \bibfield  {author} {\bibinfo {author} {\bibfnamefont {T.}~\bibnamefont
  {Barthel}}, \bibinfo {author} {\bibfnamefont {U.}~\bibnamefont
  {Schollw\"ock}}, \ and\ \bibinfo {author} {\bibfnamefont {S.~R.}\
  \bibnamefont {White}},\ }\href {\doibase 10.1103/PhysRevB.79.245101}
  {\bibfield  {journal} {\bibinfo  {journal} {Phys. Rev. B}\ }\textbf {\bibinfo
  {volume} {79}},\ \bibinfo {pages} {245101} (\bibinfo {year}
  {2009})}\BibitemShut {NoStop}%
\bibitem [{\citenamefont {Barthel}(2013)}]{Barthel_2013}%
  \BibitemOpen
  \bibfield  {author} {\bibinfo {author} {\bibfnamefont {T.}~\bibnamefont
  {Barthel}},\ }\href {\doibase 10.1088/1367-2630/15/7/073010} {\bibfield
  {journal} {\bibinfo  {journal} {New Journal of Physics}\ }\textbf {\bibinfo
  {volume} {15}},\ \bibinfo {pages} {073010} (\bibinfo {year}
  {2013})}\BibitemShut {NoStop}%
\bibitem [{\citenamefont {Kestin}\ and\ \citenamefont
  {Giamarchi}(2019)}]{Noam_Kestin_2019}%
  \BibitemOpen
  \bibfield  {author} {\bibinfo {author} {\bibfnamefont {N.}~\bibnamefont
  {Kestin}}\ and\ \bibinfo {author} {\bibfnamefont {T.}~\bibnamefont
  {Giamarchi}},\ }\href {\doibase 10.1103/PhysRevB.99.195121} {\bibfield
  {journal} {\bibinfo  {journal} {Phys. Rev. B}\ }\textbf {\bibinfo {volume}
  {99}},\ \bibinfo {pages} {195121} (\bibinfo {year} {2019})}\BibitemShut
  {NoStop}%
\bibitem [{\citenamefont {Verstraete}\ \emph {et~al.}(2004)\citenamefont
  {Verstraete}, \citenamefont {Garc\'{\i}a-Ripoll},\ and\ \citenamefont
  {Cirac}}]{Verstraete_Cirac_2004}%
  \BibitemOpen
  \bibfield  {author} {\bibinfo {author} {\bibfnamefont {F.}~\bibnamefont
  {Verstraete}}, \bibinfo {author} {\bibfnamefont {J.~J.}\ \bibnamefont
  {Garc\'{\i}a-Ripoll}}, \ and\ \bibinfo {author} {\bibfnamefont {J.~I.}\
  \bibnamefont {Cirac}},\ }\href {\doibase 10.1103/PhysRevLett.93.207204}
  {\bibfield  {journal} {\bibinfo  {journal} {Phys. Rev. Lett.}\ }\textbf
  {\bibinfo {volume} {93}},\ \bibinfo {pages} {207204} (\bibinfo {year}
  {2004})}\BibitemShut {NoStop}%
\bibitem [{\citenamefont {Schollwöck}(2011)}]{Schollwock_2011}%
  \BibitemOpen
  \bibfield  {author} {\bibinfo {author} {\bibfnamefont {U.}~\bibnamefont
  {Schollwöck}},\ }\href {\doibase https://doi.org/10.1016/j.aop.2010.09.012}
  {\bibfield  {journal} {\bibinfo  {journal} {Annals of Physics}\ }\textbf
  {\bibinfo {volume} {326}},\ \bibinfo {pages} {96} (\bibinfo {year} {2011})},\
  \bibinfo {note} {january 2011 Special Issue}\BibitemShut {NoStop}%
\bibitem [{\citenamefont {Paeckel}\ \emph {et~al.}(2019)\citenamefont
  {Paeckel}, \citenamefont {Köhler}, \citenamefont {Swoboda}, \citenamefont
  {Manmana}, \citenamefont {Schollwöck},\ and\ \citenamefont
  {Hubig}}]{Paeckel_2019}%
  \BibitemOpen
  \bibfield  {author} {\bibinfo {author} {\bibfnamefont {S.}~\bibnamefont
  {Paeckel}}, \bibinfo {author} {\bibfnamefont {T.}~\bibnamefont {Köhler}},
  \bibinfo {author} {\bibfnamefont {A.}~\bibnamefont {Swoboda}}, \bibinfo
  {author} {\bibfnamefont {S.~R.}\ \bibnamefont {Manmana}}, \bibinfo {author}
  {\bibfnamefont {U.}~\bibnamefont {Schollwöck}}, \ and\ \bibinfo {author}
  {\bibfnamefont {C.}~\bibnamefont {Hubig}},\ }\href {\doibase
  https://doi.org/10.1016/j.aop.2019.167998} {\bibfield  {journal} {\bibinfo
  {journal} {Annals of Physics}\ }\textbf {\bibinfo {volume} {411}},\ \bibinfo
  {pages} {167998} (\bibinfo {year} {2019})}\BibitemShut {NoStop}%
\bibitem [{\citenamefont {White}\ and\ \citenamefont
  {Feiguin}(2004)}]{Feiguin_White_2004}%
  \BibitemOpen
  \bibfield  {author} {\bibinfo {author} {\bibfnamefont {S.~R.}\ \bibnamefont
  {White}}\ and\ \bibinfo {author} {\bibfnamefont {A.~E.}\ \bibnamefont
  {Feiguin}},\ }\href {\doibase 10.1103/PhysRevLett.93.076401} {\bibfield
  {journal} {\bibinfo  {journal} {Phys. Rev. Lett.}\ }\textbf {\bibinfo
  {volume} {93}},\ \bibinfo {pages} {076401} (\bibinfo {year}
  {2004})}\BibitemShut {NoStop}%
\bibitem [{\citenamefont {Bouillot}\ \emph {et~al.}(2011)\citenamefont
  {Bouillot}, \citenamefont {Kollath}, \citenamefont {L\"auchli}, \citenamefont
  {Zvonarev}, \citenamefont {Thielemann}, \citenamefont {R\"uegg},
  \citenamefont {Orignac}, \citenamefont {Citro}, \citenamefont
  {Klanj\ifmmode~\check{s}\else \v{s}\fi{}ek}, \citenamefont {Berthier},
  \citenamefont {Horvati\ifmmode~\acute{c}\else \'{c}\fi{}},\ and\
  \citenamefont {Giamarchi}}]{Bouillot_2011}%
  \BibitemOpen
  \bibfield  {author} {\bibinfo {author} {\bibfnamefont {P.}~\bibnamefont
  {Bouillot}}, \bibinfo {author} {\bibfnamefont {C.}~\bibnamefont {Kollath}},
  \bibinfo {author} {\bibfnamefont {A.~M.}\ \bibnamefont {L\"auchli}}, \bibinfo
  {author} {\bibfnamefont {M.}~\bibnamefont {Zvonarev}}, \bibinfo {author}
  {\bibfnamefont {B.}~\bibnamefont {Thielemann}}, \bibinfo {author}
  {\bibfnamefont {C.}~\bibnamefont {R\"uegg}}, \bibinfo {author} {\bibfnamefont
  {E.}~\bibnamefont {Orignac}}, \bibinfo {author} {\bibfnamefont
  {R.}~\bibnamefont {Citro}}, \bibinfo {author} {\bibfnamefont
  {M.}~\bibnamefont {Klanj\ifmmode~\check{s}\else \v{s}\fi{}ek}}, \bibinfo
  {author} {\bibfnamefont {C.}~\bibnamefont {Berthier}}, \bibinfo {author}
  {\bibfnamefont {M.}~\bibnamefont {Horvati\ifmmode~\acute{c}\else
  \'{c}\fi{}}}, \ and\ \bibinfo {author} {\bibfnamefont {T.}~\bibnamefont
  {Giamarchi}},\ }\href {\doibase 10.1103/PhysRevB.83.054407} {\bibfield
  {journal} {\bibinfo  {journal} {Phys. Rev. B}\ }\textbf {\bibinfo {volume}
  {83}},\ \bibinfo {pages} {054407} (\bibinfo {year} {2011})}\BibitemShut
  {NoStop}%
\bibitem [{\citenamefont {Fukuda}\ and\ \citenamefont
  {Wortis}(1963)}]{Fukuda_Wortis_1963}%
  \BibitemOpen
  \bibfield  {author} {\bibinfo {author} {\bibfnamefont {N.}~\bibnamefont
  {Fukuda}}\ and\ \bibinfo {author} {\bibfnamefont {M.}~\bibnamefont
  {Wortis}},\ }\href {\doibase https://doi.org/10.1016/0022-3697(63)90115-X}
  {\bibfield  {journal} {\bibinfo  {journal} {Journal of Physics and Chemistry
  of Solids}\ }\textbf {\bibinfo {volume} {24}},\ \bibinfo {pages} {1675 }
  (\bibinfo {year} {1963})}\BibitemShut {NoStop}%
\bibitem [{\citenamefont {Takahashi}(1986)}]{Takahashi_1986}%
  \BibitemOpen
  \bibfield  {author} {\bibinfo {author} {\bibfnamefont {M.}~\bibnamefont
  {Takahashi}},\ }\href {https://doi.org/10.1143/PTPS.87.233} {\bibfield
  {journal} {\bibinfo  {journal} {Progress of Theoretical Physics Supplement}\
  }\textbf {\bibinfo {volume} {87}},\ \bibinfo {pages} {233} (\bibinfo {year}
  {1986})}\BibitemShut {NoStop}%
\bibitem [{\citenamefont {Takahashi}(1971)}]{Takahashi_1971}%
  \BibitemOpen
  \bibfield  {author} {\bibinfo {author} {\bibfnamefont {M.}~\bibnamefont
  {Takahashi}},\ }\href {\doibase 10.1143/PTP.46.401} {\bibfield  {journal}
  {\bibinfo  {journal} {Progress of Theoretical Physics}\ }\textbf {\bibinfo
  {volume} {46}},\ \bibinfo {pages} {401} (\bibinfo {year} {1971})}\BibitemShut
  {NoStop}%
\bibitem [{\citenamefont {Wang}\ and\ \citenamefont
  {Landau}(2001)}]{Wang_Landau_2001}%
  \BibitemOpen
  \bibfield  {author} {\bibinfo {author} {\bibfnamefont {F.}~\bibnamefont
  {Wang}}\ and\ \bibinfo {author} {\bibfnamefont {D.~P.}\ \bibnamefont
  {Landau}},\ }\href {\doibase 10.1103/PhysRevE.64.056101} {\bibfield
  {journal} {\bibinfo  {journal} {Phys. Rev. E}\ }\textbf {\bibinfo {volume}
  {64}},\ \bibinfo {pages} {056101} (\bibinfo {year} {2001})}\BibitemShut
  {NoStop}%
\bibitem [{\citenamefont {Troyer}\ \emph {et~al.}(2003)\citenamefont {Troyer},
  \citenamefont {Wessel},\ and\ \citenamefont {Alet}}]{ALPS_Troyer_2003}%
  \BibitemOpen
  \bibfield  {author} {\bibinfo {author} {\bibfnamefont {M.}~\bibnamefont
  {Troyer}}, \bibinfo {author} {\bibfnamefont {S.}~\bibnamefont {Wessel}}, \
  and\ \bibinfo {author} {\bibfnamefont {F.}~\bibnamefont {Alet}},\ }\href
  {\doibase 10.1103/PhysRevLett.90.120201} {\bibfield  {journal} {\bibinfo
  {journal} {Phys. Rev. Lett.}\ }\textbf {\bibinfo {volume} {90}},\ \bibinfo
  {pages} {120201} (\bibinfo {year} {2003})}\BibitemShut {NoStop}%
\bibitem [{\citenamefont {Bauer}\ \emph {et~al.}(2011)\citenamefont {Bauer},
  \citenamefont {Carr}, \citenamefont {Evertz}, \citenamefont {Feiguin},
  \citenamefont {Freire}, \citenamefont {Fuchs}, \citenamefont {Gamper},
  \citenamefont {Gukelberger}, \citenamefont {Gull}, \citenamefont {Guertler},
  \citenamefont {Hehn}, \citenamefont {Igarashi}, \citenamefont {Isakov},
  \citenamefont {Koop}, \citenamefont {Ma}, \citenamefont {Mates},
  \citenamefont {Matsuo}, \citenamefont {Parcollet}, \citenamefont
  {Paw{\l}owski}, \citenamefont {Picon}, \citenamefont {Pollet}, \citenamefont
  {Santos}, \citenamefont {Scarola}, \citenamefont {Schollwöck}, \citenamefont
  {Silva}, \citenamefont {Surer}, \citenamefont {Todo}, \citenamefont {Trebst},
  \citenamefont {Troyer}, \citenamefont {Wall}, \citenamefont {Werner},\ and\
  \citenamefont {Wessel}}]{Bauer_2011}%
  \BibitemOpen
  \bibfield  {author} {\bibinfo {author} {\bibfnamefont {B.}~\bibnamefont
  {Bauer}}, \bibinfo {author} {\bibfnamefont {L.~D.}\ \bibnamefont {Carr}},
  \bibinfo {author} {\bibfnamefont {H.~G.}\ \bibnamefont {Evertz}}, \bibinfo
  {author} {\bibfnamefont {A.}~\bibnamefont {Feiguin}}, \bibinfo {author}
  {\bibfnamefont {J.}~\bibnamefont {Freire}}, \bibinfo {author} {\bibfnamefont
  {S.}~\bibnamefont {Fuchs}}, \bibinfo {author} {\bibfnamefont
  {L.}~\bibnamefont {Gamper}}, \bibinfo {author} {\bibfnamefont
  {J.}~\bibnamefont {Gukelberger}}, \bibinfo {author} {\bibfnamefont
  {E.}~\bibnamefont {Gull}}, \bibinfo {author} {\bibfnamefont {S.}~\bibnamefont
  {Guertler}}, \bibinfo {author} {\bibfnamefont {A.}~\bibnamefont {Hehn}},
  \bibinfo {author} {\bibfnamefont {R.}~\bibnamefont {Igarashi}}, \bibinfo
  {author} {\bibfnamefont {S.~V.}\ \bibnamefont {Isakov}}, \bibinfo {author}
  {\bibfnamefont {D.}~\bibnamefont {Koop}}, \bibinfo {author} {\bibfnamefont
  {P.~N.}\ \bibnamefont {Ma}}, \bibinfo {author} {\bibfnamefont
  {P.}~\bibnamefont {Mates}}, \bibinfo {author} {\bibfnamefont
  {H.}~\bibnamefont {Matsuo}}, \bibinfo {author} {\bibfnamefont
  {O.}~\bibnamefont {Parcollet}}, \bibinfo {author} {\bibfnamefont
  {G.}~\bibnamefont {Paw{\l}owski}}, \bibinfo {author} {\bibfnamefont {J.~D.}\
  \bibnamefont {Picon}}, \bibinfo {author} {\bibfnamefont {L.}~\bibnamefont
  {Pollet}}, \bibinfo {author} {\bibfnamefont {E.}~\bibnamefont {Santos}},
  \bibinfo {author} {\bibfnamefont {V.~W.}\ \bibnamefont {Scarola}}, \bibinfo
  {author} {\bibfnamefont {U.}~\bibnamefont {Schollwöck}}, \bibinfo {author}
  {\bibfnamefont {C.}~\bibnamefont {Silva}}, \bibinfo {author} {\bibfnamefont
  {B.}~\bibnamefont {Surer}}, \bibinfo {author} {\bibfnamefont
  {S.}~\bibnamefont {Todo}}, \bibinfo {author} {\bibfnamefont {S.}~\bibnamefont
  {Trebst}}, \bibinfo {author} {\bibfnamefont {M.}~\bibnamefont {Troyer}},
  \bibinfo {author} {\bibfnamefont {M.~L.}\ \bibnamefont {Wall}}, \bibinfo
  {author} {\bibfnamefont {P.}~\bibnamefont {Werner}}, \ and\ \bibinfo {author}
  {\bibfnamefont {S.}~\bibnamefont {Wessel}},\ }\href {\doibase
  10.1088/1742-5468/2011/05/p05001} {\bibfield  {journal} {\bibinfo  {journal}
  {Journal of Statistical Mechanics: Theory and Experiment}\ }\textbf {\bibinfo
  {volume} {2011}},\ \bibinfo {pages} {P05001} (\bibinfo {year}
  {2011})}\BibitemShut {NoStop}%
\bibitem [{\citenamefont {Halpin-Healy}(1989)}]{Timothy}%
  \BibitemOpen
  \bibfield  {author} {\bibinfo {author} {\bibfnamefont {T.}~\bibnamefont
  {Halpin-Healy}},\ }\href {\doibase 10.1103/PhysRevB.40.772} {\bibfield
  {journal} {\bibinfo  {journal} {Phys. Rev. B}\ }\textbf {\bibinfo {volume}
  {40}},\ \bibinfo {pages} {772} (\bibinfo {year} {1989})}\BibitemShut
  {NoStop}%
\bibitem [{\citenamefont {Tonegawa}(1970)}]{Tonegawa_1970}%
  \BibitemOpen
  \bibfield  {author} {\bibinfo {author} {\bibfnamefont {T.}~\bibnamefont
  {Tonegawa}},\ }\href {https://doi.org/10.1143/PTPS.46.61} {\bibfield
  {journal} {\bibinfo  {journal} {Progress of Theoretical Physics Supplement}\
  }\textbf {\bibinfo {volume} {46}},\ \bibinfo {pages} {61} (\bibinfo {year}
  {1970})}\BibitemShut {NoStop}%
\bibitem [{\citenamefont {Papanicolaou}(1988)}]{Papanicolaou_1988}%
  \BibitemOpen
  \bibfield  {author} {\bibinfo {author} {\bibfnamefont {N.}~\bibnamefont
  {Papanicolaou}},\ }\href {\doibase
  https://doi.org/10.1016/0550-3213(88)90073-9} {\bibfield  {journal} {\bibinfo
   {journal} {Nuclear Physics B}\ }\textbf {\bibinfo {volume} {305}},\ \bibinfo
  {pages} {367 } (\bibinfo {year} {1988})}\BibitemShut {NoStop}%
\bibitem [{\citenamefont {Manmana}\ \emph {et~al.}(2011)\citenamefont
  {Manmana}, \citenamefont {L\"auchli}, \citenamefont {Essler},\ and\
  \citenamefont {Mila}}]{Salvatore_2011}%
  \BibitemOpen
  \bibfield  {author} {\bibinfo {author} {\bibfnamefont {S.~R.}\ \bibnamefont
  {Manmana}}, \bibinfo {author} {\bibfnamefont {A.~M.}\ \bibnamefont
  {L\"auchli}}, \bibinfo {author} {\bibfnamefont {F.~H.~L.}\ \bibnamefont
  {Essler}}, \ and\ \bibinfo {author} {\bibfnamefont {F.}~\bibnamefont
  {Mila}},\ }\href {\doibase 10.1103/PhysRevB.83.184433} {\bibfield  {journal}
  {\bibinfo  {journal} {Phys. Rev. B}\ }\textbf {\bibinfo {volume} {83}},\
  \bibinfo {pages} {184433} (\bibinfo {year} {2011})}\BibitemShut {NoStop}%
\bibitem [{\citenamefont {Aghahosseini}\ and\ \citenamefont
  {Parkinson}(1978)}]{Aghahosseini_Parkinson_1978}%
  \BibitemOpen
  \bibfield  {author} {\bibinfo {author} {\bibfnamefont {H.}~\bibnamefont
  {Aghahosseini}}\ and\ \bibinfo {author} {\bibfnamefont {J.~B.}\ \bibnamefont
  {Parkinson}},\ }\href@noop {} {\bibfield  {journal} {\bibinfo  {journal}
  {Journal of Physics C: Solid State Physics}\ }\textbf {\bibinfo {volume}
  {11}},\ \bibinfo {pages} {461} (\bibinfo {year} {1978})}\BibitemShut
  {NoStop}%
\bibitem [{\citenamefont {Mila}\ and\ \citenamefont
  {Zhang}(2000)}]{Mila_Zhang_2000}%
  \BibitemOpen
  \bibfield  {author} {\bibinfo {author} {\bibfnamefont {F.}~\bibnamefont
  {Mila}}\ and\ \bibinfo {author} {\bibfnamefont {F.-C.}\ \bibnamefont
  {Zhang}},\ }\href {https://doi.org/10.1007/s100510070242} {\bibfield
  {journal} {\bibinfo  {journal} {The European Physical Journal B - Condensed
  Matter and Complex Systems}\ }\textbf {\bibinfo {volume} {16}} (\bibinfo
  {year} {2000})}\BibitemShut {NoStop}%
\bibitem [{\citenamefont {Silberglitt}\ and\ \citenamefont
  {Torrance}(1970)}]{Silberglitt_Torrance_1970}%
  \BibitemOpen
  \bibfield  {author} {\bibinfo {author} {\bibfnamefont {R.}~\bibnamefont
  {Silberglitt}}\ and\ \bibinfo {author} {\bibfnamefont {J.~B.}\ \bibnamefont
  {Torrance}},\ }\href {\doibase 10.1103/PhysRevB.2.772} {\bibfield  {journal}
  {\bibinfo  {journal} {Phys. Rev. B}\ }\textbf {\bibinfo {volume} {2}},\
  \bibinfo {pages} {772} (\bibinfo {year} {1970})}\BibitemShut {NoStop}%
\bibitem [{\citenamefont {Papanicolaou}\ and\ \citenamefont
  {Psaltakis}(1987)}]{Papanicolaou_1987}%
  \BibitemOpen
  \bibfield  {author} {\bibinfo {author} {\bibfnamefont {N.}~\bibnamefont
  {Papanicolaou}}\ and\ \bibinfo {author} {\bibfnamefont {G.~C.}\ \bibnamefont
  {Psaltakis}},\ }\href {\doibase 10.1103/PhysRevB.35.342} {\bibfield
  {journal} {\bibinfo  {journal} {Phys. Rev. B}\ }\textbf {\bibinfo {volume}
  {35}},\ \bibinfo {pages} {342} (\bibinfo {year} {1987})}\BibitemShut
  {NoStop}%
\bibitem [{\citenamefont {Inami}\ \emph {et~al.}(1994)\citenamefont {Inami},
  \citenamefont {Kakurai}, \citenamefont {Tanaka}, \citenamefont {Enderle},\
  and\ \citenamefont {Steiner}}]{Toshiya_Inami1994}%
  \BibitemOpen
  \bibfield  {author} {\bibinfo {author} {\bibfnamefont {T.}~\bibnamefont
  {Inami}}, \bibinfo {author} {\bibfnamefont {K.}~\bibnamefont {Kakurai}},
  \bibinfo {author} {\bibfnamefont {H.}~\bibnamefont {Tanaka}}, \bibinfo
  {author} {\bibfnamefont {M.}~\bibnamefont {Enderle}}, \ and\ \bibinfo
  {author} {\bibfnamefont {M.}~\bibnamefont {Steiner}},\ }\href {\doibase
  10.1143/jpsj.63.1530} {\bibfield  {journal} {\bibinfo  {journal} {Journal of
  the Physical Society of Japan}\ }\textbf {\bibinfo {volume} {63}},\ \bibinfo
  {pages} {1530} (\bibinfo {year} {1994})}\BibitemShut {NoStop}%
\bibitem [{\citenamefont {Elliott}\ and\ \citenamefont
  {Thorpe}(1969)}]{Elliott_1969}%
  \BibitemOpen
  \bibfield  {author} {\bibinfo {author} {\bibfnamefont {R.~J.}\ \bibnamefont
  {Elliott}}\ and\ \bibinfo {author} {\bibfnamefont {M.~F.}\ \bibnamefont
  {Thorpe}},\ }\href {\doibase 10.1088/0022-3719/2/9/312} {\bibfield  {journal}
  {\bibinfo  {journal} {Journal of Physics C: Solid State Physics}\ }\textbf
  {\bibinfo {volume} {2}},\ \bibinfo {pages} {1630} (\bibinfo {year}
  {1969})}\BibitemShut {NoStop}%
\bibitem [{\citenamefont {Headings}\ \emph {et~al.}(2010)\citenamefont
  {Headings}, \citenamefont {Hayden}, \citenamefont {Coldea},\ and\
  \citenamefont {Perring}}]{Headings_2010}%
  \BibitemOpen
  \bibfield  {author} {\bibinfo {author} {\bibfnamefont {N.~S.}\ \bibnamefont
  {Headings}}, \bibinfo {author} {\bibfnamefont {S.~M.}\ \bibnamefont
  {Hayden}}, \bibinfo {author} {\bibfnamefont {R.}~\bibnamefont {Coldea}}, \
  and\ \bibinfo {author} {\bibfnamefont {T.~G.}\ \bibnamefont {Perring}},\
  }\href {\doibase 10.1103/PhysRevLett.105.247001} {\bibfield  {journal}
  {\bibinfo  {journal} {Phys. Rev. Lett.}\ }\textbf {\bibinfo {volume} {105}},\
  \bibinfo {pages} {247001} (\bibinfo {year} {2010})}\BibitemShut {NoStop}%
\bibitem [{\citenamefont {Mourigal}\ \emph {et~al.}(2013)\citenamefont
  {Mourigal}, \citenamefont {Enderle}, \citenamefont {Kl{\"o}pperpieper},
  \citenamefont {Caux}, \citenamefont {Stunault},\ and\ \citenamefont
  {R{\o}nnow}}]{Mourigal2013}%
  \BibitemOpen
  \bibfield  {author} {\bibinfo {author} {\bibfnamefont {M.}~\bibnamefont
  {Mourigal}}, \bibinfo {author} {\bibfnamefont {M.}~\bibnamefont {Enderle}},
  \bibinfo {author} {\bibfnamefont {A.}~\bibnamefont {Kl{\"o}pperpieper}},
  \bibinfo {author} {\bibfnamefont {J.-S.}\ \bibnamefont {Caux}}, \bibinfo
  {author} {\bibfnamefont {A.}~\bibnamefont {Stunault}}, \ and\ \bibinfo
  {author} {\bibfnamefont {H.~M.}\ \bibnamefont {R{\o}nnow}},\ }\href {\doibase
  10.1038/nphys2652} {\bibfield  {journal} {\bibinfo  {journal} {Nature
  Physics}\ }\textbf {\bibinfo {volume} {9}},\ \bibinfo {pages} {435} (\bibinfo
  {year} {2013})}\BibitemShut {NoStop}%
\bibitem [{\citenamefont {Jeske}\ \emph {et~al.}(2018)\citenamefont {Jeske},
  \citenamefont {Rivas}, \citenamefont {Ahmed}, \citenamefont
  {Martin-Delgado},\ and\ \citenamefont {Cole}}]{Jeske_2018}%
  \BibitemOpen
  \bibfield  {author} {\bibinfo {author} {\bibfnamefont {J.}~\bibnamefont
  {Jeske}}, \bibinfo {author} {\bibfnamefont {{\'{A}}.}~\bibnamefont {Rivas}},
  \bibinfo {author} {\bibfnamefont {M.~H.}\ \bibnamefont {Ahmed}}, \bibinfo
  {author} {\bibfnamefont {M.~A.}\ \bibnamefont {Martin-Delgado}}, \ and\
  \bibinfo {author} {\bibfnamefont {J.~H.}\ \bibnamefont {Cole}},\ }\href
  {\doibase 10.1088/1367-2630/aadecf} {\bibfield  {journal} {\bibinfo
  {journal} {New Journal of Physics}\ }\textbf {\bibinfo {volume} {20}},\
  \bibinfo {pages} {093017} (\bibinfo {year} {2018})}\BibitemShut {NoStop}%
\bibitem [{\citenamefont {Keselman}\ \emph {et~al.}(2020)\citenamefont
  {Keselman}, \citenamefont {Balents},\ and\ \citenamefont
  {Starykh}}]{Keselman_2020}%
  \BibitemOpen
  \bibfield  {author} {\bibinfo {author} {\bibfnamefont {A.}~\bibnamefont
  {Keselman}}, \bibinfo {author} {\bibfnamefont {L.}~\bibnamefont {Balents}}, \
  and\ \bibinfo {author} {\bibfnamefont {O.~A.}\ \bibnamefont {Starykh}},\
  }\href {\doibase 10.1103/PhysRevLett.125.187201} {\bibfield  {journal}
  {\bibinfo  {journal} {Phys. Rev. Lett.}\ }\textbf {\bibinfo {volume} {125}},\
  \bibinfo {pages} {187201} (\bibinfo {year} {2020})}\BibitemShut {NoStop}%
\end{thebibliography}%
\end{document}